\documentclass[trackchanges,twocolumn]{aastex701}
\usepackage{placeins}
\usepackage{cancel}

\newcommand{\Mchar}{10^{11.5}\,M_\odot}
\usepackage{amsmath,amssymb}
\begin{document}

\title{Primordial Black Holes as Cosmic Architects: \\Imprints of an extended mass function}

\author[orcid=0009-0009-6060-9540]{Alberto Magaraggia}
\affiliation{Department of Physics, University of Miami, Coral Gables, FL 33124, USA}
\email[]{axm8568@miami.edu}
\author[orcid=0000-0002-1697-186X]{Nico Cappelluti}
\email[show]{ncappelluti@miami.edu}
\affiliation{Department of Physics, University of Miami, Coral Gables, FL 33124, USA}
\author[orcid=0000-0002-0797-0646]{G\"unther Hasinger}
\email[]{guenther.hasinger@dzastro.de}
\affiliation{TU Dresden, Institute of Nuclear and Particle Physics, 01062 Dresden, Germany}
\affiliation{Deutsches Zentrum f\"ur Astrophysik, Postplatz 1, 02826 G\"orlitz, Germany}
\author[0009-0008-2685-3497]{Ma\"el Gonin}
\email[]{mael.gonin@tu-dresden.de}
\affiliation{TU Dresden, Institute of Nuclear and Particle Physics, 01062 Dresden, Germany}
\affiliation{Deutsches Zentrum f\"ur Astrophysik, Postplatz 1, 02826 G\"orlitz, Germany}
\author[orcid=0000-0002-5554-8896]{Priyamvada Natarajan}
\email[]{priyamvada.natarajan@yale.edu}
\affiliation{Black Hole Initiative at Harvard University, 20 Garden Street, Cambridge, MA 02138, USA}
\affiliation{Department of Astronomy, Yale University, Kline Tower, 266 Whitney Avenue, New Haven, CT 06511, USA}
\affiliation{Department of Physics, Yale University, P.O. Box 208121, New Haven, CT 06520, USA}

\begin{abstract} 
Primordial black holes (PBHs) remain compelling dark-matter candidates and possible relics of early-Universe physics, with observable imprints across both cosmological structure formation and gravitational-wave astronomy. We investigate the consequences of a PBH population described by an extended, physically motivated mass function shaped by features in the thermal history of the early Universe. We compute the associated Poisson contribution to the matter power spectrum and propagate its effects to the halo mass function, the cosmic star-formation rate density, and the reionization history. The PBH component produces a moderate enhancement of small-scale power, leading to earlier low-mass halo formation and a correspondingly mild increase in the high-redshift star-formation rate density while remaining consistent with current constraints on the Thomson optical depth. We then evaluate the merger-rate distribution expected for current and future gravitational-wave detectors. In the stellar-mass regime, the predicted PBH binary population occupies the region probed by LIGO--Virgo--KAGRA and can contribute to the observed compact-object merger population. For LISA, the detectable signal is dominated by intermediate-mass binaries with component masses of $10^3$--$10^4\,M_\odot$, with cumulative rates of order a few events per year extending to redshifts as high as $z \sim 50$. The detection of such high-redshift intermediate-mass black hole mergers would constitute a distinctive signature of a primordial origin, providing evidence that PBHs contribute non-negligibly to the dark matter budget and may have served as early seeds for the supermassive black holes observed at later cosmic epochs. 
\end{abstract}


\section{Introduction}
Primordial black holes (PBHs), first proposed more than five decades ago \citep{hawking71,carr74,1975Natur.253..251C}, are black holes that may have formed in the early Universe from the collapse of large primordial density fluctuations. Unlike astrophysical black holes, whose formation is tied to stellar evolution \citep{MIRABEL20171}, PBHs can span an enormous mass range, set by the horizon mass at formation and by the underlying physical mechanism \citep{MaximYuKhlopov_2010}. In principle, their masses may range from near the Planck scale, $M \sim 10^{-5}\;\mathrm{g}$, to masses comparable to those of supermassive black holes (SMBHs) \citep{Carr2025}. This wide dynamic range makes PBHs a distinctive probe of the early Universe and a potentially important ingredient in addressing several outstanding problems in the standard $\Lambda$CDM cosmological framework.

The nature of dark matter (DM) remains one of the central unresolved questions in cosmology. Despite extensive experimental and observational searches for direct particle evidence, dark matter has so far been detected only through its gravitational effects. A broad set of candidates has been proposed, including light bosons, neutrinos, weakly interacting massive particles, modified-gravity scenarios, weak-scale relics, and compact objects such as PBHs \citep{cirelli2025darkmatter}. PBHs are especially compelling because, while their formation requires special initial conditions, they do not necessarily require the introduction of new particle species \citep{PhysRevD.98.063005}. At the same time, they can retain imprints of physics beyond the Standard Model through their formation history and mass spectrum \citep{sym16111487}. Depending on their abundance,
$f_{\mathrm{PBH}} \equiv \frac{\Omega_{\mathrm{PBH}}}{\Omega_{\mathrm{DM}}}$, PBHs may constitute a subdominant, significant, or even a potentially dominant fraction of the dark matter. PBHs have also been proposed as natural seeds for the SMBHs observed at redshifts $z \gtrsim 6$ \citep{Mack_2007}. The existence of such massive black holes within the first billion years of cosmic time remains difficult to explain through Eddington-limited growth from stellar-mass remnants alone \citep{SMBHrefId0}. This tension has motivated a range of massive-black-hole seeding channels, including direct collapse \citep{Natarajan+2017}, dense stellar dynamics, and primordial formation. Gravitational-wave (GW) observations provide an independent and complementary probe of these possibilities. 
The Laser Interferometer Space Antenna (LISA)\footnote{\url{https://www.lisamission.org/}}, a planned ESA-led space-based GW observatory, will be sensitive to black-hole mergers with masses $\sim 10^{3}$--$10^{8}\,M_\odot$ out to very high redshift \citep{bellovary2019intermediatemassblackholes}. By measuring the masses, redshifts, and merger rates of early black-hole binaries, LISA will directly constrain the origin, abundance, and growth history of massive black-hole seeds \citep{Ricarte+2018}.

The stellar-mass regime is already being probed by the LIGO\footnote{\url{https://ligo.org/}}--Virgo\footnote{\url{https://www.virgo-gw.eu/}}--KAGRA\footnote{\url{https://gwcenter.icrr.u-tokyo.ac.jp/}} (LVK) network. Several studies have noted that a subset of the compact-binary mergers detected by LVK may be consistent with a primordial origin. Early work showed that the masses and merger rates of some LIGO black-hole binaries could be compatible with PBHs contributing to the dark matter \citep{Bird_2016}. More recently, searches for subsolar-mass compact objects have reported candidate events with component masses below the range expected from standard stellar evolution; if confirmed, such systems would provide a particularly clean signature of a primordial origin \citep{PhysRevD.106.123526,Prunier_2024,Magaraggia2026,haque2026primordialblackholeinterpretation}. Current and forthcoming GW data will therefore provide increasingly powerful constraints on PBH abundance, mass distribution, and binary formation channels, while helping to distinguish PBHs from astrophysical black holes and other exotic compact objects \citep{PhysRevD.109.124063Crescimbeni,Crescimbeni:2024qrq}.

A major theoretical uncertainty is the PBH mass function. While monochromatic mass functions are often adopted for simplicity, realistic formation scenarios generically produce extended distributions rather than a single characteristic mass. Such distributions may be described phenomenologically, for example by log-normal or power-law forms \citep{PhysRevD.96.023514,Bellomo2018-fz}, or derived from specific early-Universe formation mechanisms \citep{Hasinger2020,PhysRevD.106.123526,Pritchard:2024vix}. In particular, features in the thermal history of the Universe, including changes in the equation of state associated with particle thresholds, annihilation epochs, and phase transitions, can imprint characteristic structure on the PBH mass spectrum.

PBH abundances are constrained by a wide range of observations, typically through the non-detection of mass-dependent astrophysical and cosmological signatures \citep{Carr2025}. Although these constraints are stringent, non-negligible windows remain open. Moreover, most published limits assume monochromatic PBH mass functions and therefore cannot be applied directly to broad extended distributions \citep{Carr2025}. For an extended mass function, each observable must be convolved with the full distribution, and the corresponding constraint must be re-derived or mapped to an appropriate effective monochromatic mass \citep{PhysRevD.96.023514,Bellomo2018-fz}. This procedure is tractable for sufficiently narrow distributions but becomes technically challenging, and in some cases ill-defined, for very broad mass functions \citep{Bellomo2018-fz}. Direct comparison between extended PBH models and monochromatic bounds must therefore be interpreted with care. This caveat is particularly relevant for microlensing constraints, which are among the most commonly quoted bounds on stellar-mass PBHs. In addition to the difficulty of translating monochromatic limits to an extended mass function, the interpretation of microlensing surveys depends on assumptions about the survey detection efficiency, source populations, self-lensing backgrounds, and the Galactic halo model. In this context,
\cite{10.1093/mnras/staf1826} have argued that the recent OGLE limits toward the Large Magellanic Cloud (LMC) should not be regarded as robust constraints on a stellar-mass PBH population, because the OGLE event sample falls below even the expected self-lensing background and is in tension with the earlier MACHO detections. While the status of LMC microlensing constraints remains debated, this illustrates that existing PBH bounds, especially when applied to broad mass functions, must be interpreted with caution.

In this work, we investigate the cosmological and gravitational-wave consequences of a PBH population described by an extended, physically motivated mass function. We compute the PBH contribution to the matter power spectrum and propagate its impact into the halo mass function, the cosmic star-formation-rate density, and the reionization history. We then evaluate the expected PBH binary merger rates for current and future GW observatories, focusing on the LVK-accessible stellar-mass regime and the LISA-accessible intermediate and massive-black-hole regime. Section~2 describes the adopted PBH mass function and the methodology used to model its effects on structure formation and GW signals. Section~3 presents the resulting cosmological and observational implications, including consistency with current constraints and prospects for distinguishing primordial from astrophysical black-hole populations.

Throughout this paper, we assume a flat Universe with Planck cosmological parameters \citep{Planck_Collaboration2020-ij}: $\Omega_\Lambda = 0.68885$, $\Omega_m=0.30966$, $\Omega_b = 0.04897$, $\Omega_{dm}=\Omega_m - \Omega_b$, $\sigma_8 = 0.8102$, and $h= 0.6766$. Here $\Omega_\Lambda$, $\Omega_m$, $\Omega_{dm}$, and $\Omega_b$ denote the dark-energy, total-matter, dark-matter, and baryon density parameters relative to the critical density, $\rho_c = \frac{3H_0^2}{8\pi G}$, with $H_0 = 100h\;{\rm km\,s^{-1}\,Mpc^{-1}}$, and $\sigma_8$ the rms amplitude of matter fluctuations on $8h^{-1}\,{\rm Mpc}$ scales. For each cosmic component $i$, we define its energy density as $\rho_i \equiv \Omega_i \rho_c$.

\section{Methods}

\subsection{The PBH mass function}
PBHs can provide a unique probe of the early Universe physics and its thermal history \citep{Carr2021-lr,CARR20241}. 
The cosmic equation of state (EoS) is described as the ratio between the pressure and the energy density of the primordial plasma, $w(T) \equiv {P(T) /\epsilon(T)}$.
The primordial PBH mass function can be obtained by translating the thermal history of the early Universe into a collapse probability \citep{Gonin2026} for their formation. Since PBHs form when sufficiently large overdensities re-enter the horizon during radiation domination, each temperature can be associated with a horizon mass $M_H(T)$, which is identified, in the simplest analytical treatment, with a corresponding PBH mass. The dependence on the EoS then enters through the collapse threshold $\delta_c(w)$: when the plasma becomes softer, namely when $w$ dips below the radiation value (1/3), pressure support is reduced and overdensities collapse more easily. Assuming a primordial spectrum of Gaussian density fluctuations, this gives a mass-dependent collapse fraction $\beta(M)$, which is finally converted into the PBH abundance per logarithmic mass interval, $df_{\rm PBH}/d\ln M$ shown in Figure~1.
In this way, features in the early-Universe EoS, such as QCD confinement at a pseudo-critical temperature $T_c\sim156 \rm~MeV$ \citep{HotQCD:2018pds}, as well as particle annihilation epochs, 
can soften the equation of state and enhance PBH formation at the corresponding horizon masses, giving rise to a multimodal and physically motivated PBH mass spectrum \citep{Byrnes_2018}. 

Moreover, taking into account primordial asymmetries between particles and anti-particles in the baryonic and leptonic sector appears to be of primary importance as they induce chemical potentials in the particle distribution \citep{Schwarz:2009ii,Wygas:2018otj,Middeldorf-Wygas:2020glx,Formaggio:2025nde}. 
The consideration of these asymmetries adds 5 charges to conserve:
\begin{subequations}
\label{eq:conservation_equations}
\begin{align}
    \ell_\alpha s &= n_\alpha + n_{\nu_\alpha} = n_{L_\alpha}, \label{eq:conservation_equations:a}\\
    bs &= \sum_i B_i n_i = n_B, \label{eq:conservation_equations:b}\\
    qs &= \sum_i Q_i n_i = n_Q \label{eq:conservation_equations:c}
\end{align}
\end{subequations}
Here $s$ denotes the total entropy density, $n_i$ is the net number density of particle species $i$, and $B_i$ and $Q_i$ are its baryon number and electric charge, respectively. The quantities $n_B$ and $n_Q$ are the total net baryon and electric charge number densities, while $b \equiv n_B/s$ and $q \equiv n_Q/s$ are the corresponding asymmetries normalized to the entropy density. We take $b=8.6\times10^{-11}$, as inferred from \citep{Planck_Collaboration2020-ij}, and impose electric charge neutrality, $q=0$, consistently with constraints on the electric charge of the Universe \citep{Caprini:2003gz}. For each lepton flavor $\alpha \in \{e,\mu,\tau\}$, $n_\alpha$ denotes the net charged-lepton number density and $n_{\nu_\alpha}$ the net neutrino number density, so that $n_{L_\alpha}=n_\alpha+n_{\nu_\alpha}$ is the conserved net lepton-flavor number density and $\ell_\alpha \equiv n_{L_\alpha}/s$ is the corresponding lepton-flavor asymmetry. The values of $\ell_\alpha$ are constrained by CMB and BBN observations \citep{Froustey:2024mgf,Domcke2025}. 
Recent analysis suggest the electronic neutrino asymmetry $\ell_{\nu_e}\sim10^{-2}$ \citep{Li:2024gzf}. 
However, once neutrino oscillations set in at $T\lesssim 10,\rm MeV$, flavor asymmetries are mixed and even initially large primordial asymmetries can become consistent with observational constraints. Larger primordial asymmetries are therefore allowed \citep{Froustey:2024mgf}.
Non-zero lepton flavor asymmetries prior to the onset of neutrino oscillations can significantly alter the effective relativistic degrees of freedom and the equation of state around the QCD epoch, thereby reshaping the PBH mass function \citep{Bodeker2021-hz,Gonin2026}. 
Due to electric charge conservation (Eq.~\eqref{eq:conservation_equations:c}), the chemical potential of leptons is correlated with the baryonic chemical potential. The introduction of the cosmic asymmetries $b$, $\ell_e$, $\ell_\mu$, and $\ell_\tau$ can cause the QCD transition to occur at large values of $\mu_B$.
Some studies suggest the possibility of a first-order order QCD transition powered by lepton flavor asymmetry \citep{Gao:2021nwz,Gao:2023djs,Gao:2024fhm}.
Moreover, the inclusion of chemical potentials in the leptonic distributions makes their overall contribution to the early plasma more significant. As a result, events like $\tau^+\tau^-$ annihilation, followed by the transfer of the lepton number to $\nu_\tau$, can leave a deep imprint in the cosmic EoS. When neutrinos contribute significantly to the plasma, the EoS tilts to the pure radiation value $w=1/3$, mitigating peaks in the PBH distribution.

\begin{figure}
    \centering
    \includegraphics[width=\linewidth]{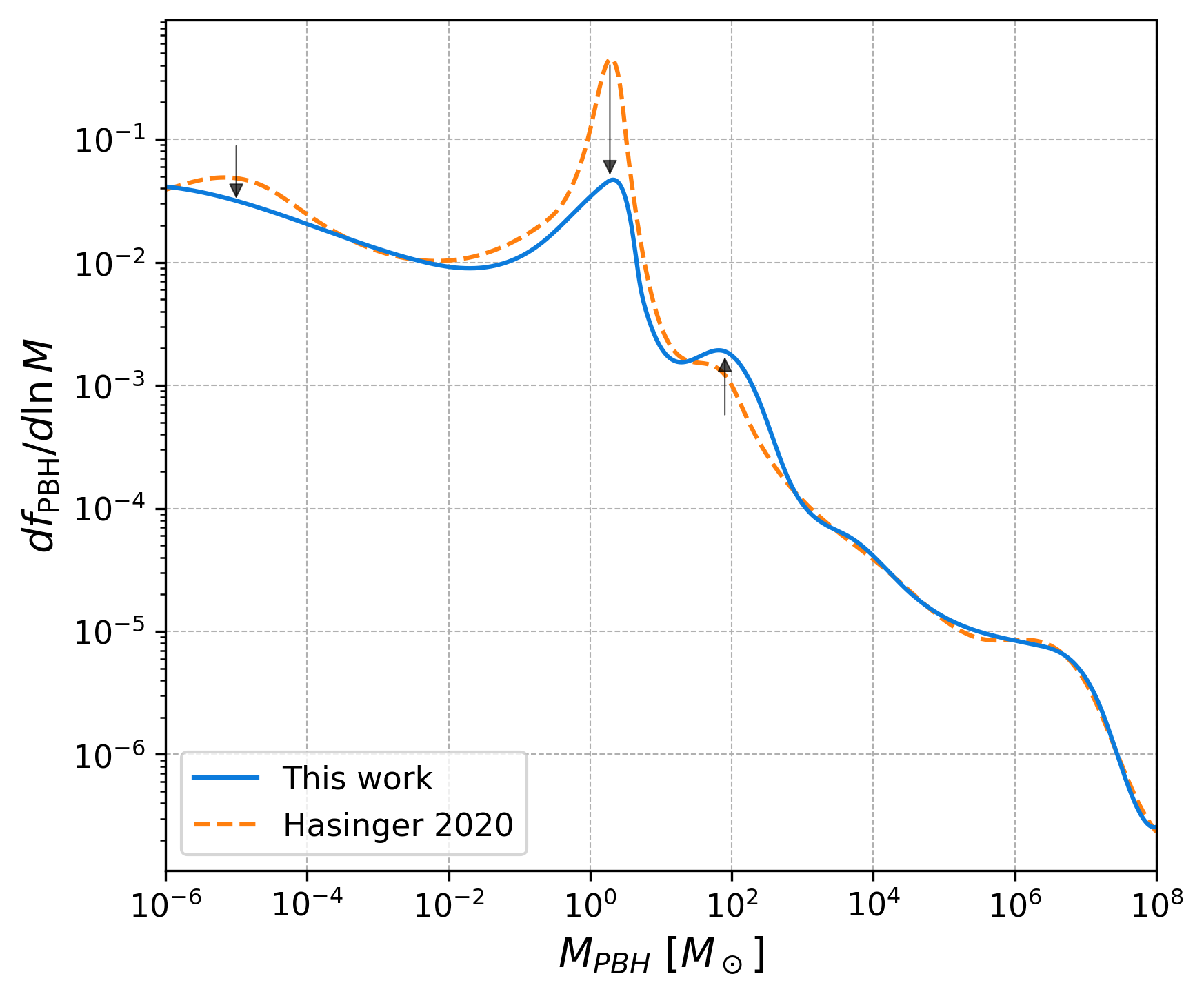}
    \caption{The PBH mass function assumed in this work is depicted with the blue, solid line. It is derived from the orange, dashed line \citep{Hasinger2020} to which a lepton flavour asymmetry has been added \citep{Bodeker2021-hz}. The lepton flavour asymmetry lowers the bumps at $10^{-5}\,M_\odot$, $1.5 \;M_{\odot}$, and it enhances the one at $\sim 100 \;M_{\odot}$.}
    \label{fig:massfunction}
\end{figure}

Following this motivation, we employ the extended PBH mass function computed by \citet{Hasinger2020} using the procedure outlined above, which was provided to us through private communication.
 In this framework, four distinct populations of primordial black holes are generated across a broad mass spectrum ($10^{-6} M_\odot < M_{PBH} < 10^{8} M_\odot$). Specifically, planetary-mass PBHs ($\sim10^{-5}\,M_\odot$) originate at the electro weak transitions; PBHs with masses near the Chandrasekhar limit ($\sim1.5\,M_\odot$) form when baryons (protons and neutrons) emerge from three-quark states; PBHs of characteristic masses around $50\,M_\odot$ arise during the formation of pions from two quarks, quickly followed by their annihilation corresponding to the disappearance of anti-quarks; and finally, supermassive PBHs ($\sim10^{7}\,M_\odot$) are produced at the epoch of $e^{+}e^{-}$ annihilation \citep{Hasinger2020}. These four populations correspond to the peaks in the PBH mass function, $\mathrm{d}f_{\rm PBH}/\mathrm{d}\ln M$, shown with a dashed orange line in Figure \ref{fig:massfunction}.

The mass function from \cite{Hasinger2020} has been modified (G. Hasinger 2025, private communication) to take into account the presence of sizeable lepton asymmetries, choosing the values $\ell_e = -8\times 10^{-2}$ and $\ell_\mu = \ell_\tau = 4\times 10^{-2}$ \citep{Bodeker2021-hz}, which yield a distinctive redistribution of PBHs formation probability across the mass range: the planetary mass PBHs at ($\sim10^{-5}\,M_\odot$) are reduced, the original peak at $\sim 1.5\;M_{\odot}$ is highly suppressed, and a small bump forms at around $100\;M_{\odot}$ (corresponding to the three black arrows of Figure \ref{fig:massfunction}). 
The resulting PBH mass function adopted in this paper is depicted with a solid blue line in Figure \ref{fig:massfunction}. We note that the PBH mass function is intrinsically sensitive to the detailed particle content of the early Universe, and any modification to the EoS -- such as the appearance of new particle species -- can leave characteristic imprints on the collapse probability and thus on the resulting PBH mass spectrum. 
For instance, the inclusion of additional degrees of freedom (e.g., beyond the Standard Model, such as the X17 scenario) alters the thermodynamic evolution and leads to measurable shifts in the PBH mass distribution \citep{Gonin2025}. In principle, a fully consistent computation of the PBH mass function would therefore require an accurate treatment of the EoS across all relevant temperature regimes, including all particle interactions. 

However, this rapidly becomes technically challenging, as the EoS acquires a highly non-trivial structure due to the interplay of phase transitions, particle production, and annihilation processes. In this work, we adopt a simplified, phenomenological approach as described above, which captures the main physically motivated features of the mass spectrum. We emphasize, nevertheless, that the true PBH mass function may differ quantitatively, and we defer a more complete treatment to future work. The electroweak phase transition is the portion of the spectrum that would require the most attention, as it could be the place where baryogenesis and leptogenesis occur \citep{Bodeker:2020ghk}. In order to remain as agnostic as possible only a slight deviation between the `Hasinger 2020' and the flavour asymmetric case is introduced. On the other hand, the difference around $M_{PBH}\sim10^7 ~M_\odot$, corresponding to $e^+e^-$ annihilation, is fairly accurate, as the introduction of chemical potential to the neutrinos does not drastically modify the EoS at $e^+e^-$ annihilation (private communication with Julien Froustey).

Throughout our analysis, we adopt a total PBH dark-matter fraction of $f_{\mathrm{PBH}}=\int \frac{df_{PBH}}{dM}\, dM = 0.338$ over the mass range of interest ($10^{-6} M_\odot < M_{PBH} < 10^{8} M_\odot$). However, the integral over the whole mass range (i.e., extending the analysis to sub-asteroidal PBHs, $M_{PBH} < 10^{-6} M_\odot $, not analyzed in this paper) yields $f_{\mathrm{PBH}}=1$, thereby exploring a scenario in which PBHs constitute the entire dark matter budget while carrying a theoretically motivated mass spectrum linked to early-Universe particle physics. 
As pointed out in the Introduction section, PBH constraints are extremely challenging to compute when broad mass functions are considered: as emphasized by \citep{Bellomo2018-fz}, the mass dependence of each observable must be properly taken into account by convolving it with the extended mass function. A convenient way to proceed is to associate the extended distribution with an “equivalent” monochromatic mass that produces the same effect for the observable in question. Only after this step can monochromatic constraints be meaningfully reinterpreted \citep{PhysRevD.96.023514,Bellomo2018-fz}. Without such a mapping, comparisons between extended mass functions and published monochromatic bounds are generally unreliable.
An attempt at a qualitative comparison between this mass function and current constraints is presented in \cite{Magaraggia2026}. We stress, however, that this exercise has been intended purely as a visualization of how the distribution relates to existing bounds, and does not involve performing the mapping to an equivalent monochromatic mass required for a meaningful reinterpretation of those constraints.

\subsection{Matter power spectrum}

The matter power spectrum (PS) quantifies the statistical distribution of matter density fluctuations as a function of spatial scale and is one of the primary tools for studying the formation and evolution of cosmic structure. It encodes how primordial perturbations, seeded during inflation, have evolved under the influence of gravity, cosmic expansion, and baryonic physics. Observationally, the power spectrum is constrained by measurements of cosmic microwave background anisotropies; galaxy clustering; weak gravitational lensing; and the Lyman-$\alpha$ forest - that together provide a consistent picture of structure formation across cosmic time \citep{PhysRevD.66.103508_PS}. 
In the standard $\Lambda$CDM framework, the shape of the matter power spectrum reflects the initial nearly scale-invariant spectrum of curvature perturbations, subsequently modified by the transfer of power through radiation–matter equality and the growth of density perturbations in the matter-dominated era. 
Deviations from the expected form of the matter power spectrum can signal new physics beyond the standard model of cosmology. In particular, a population of primordial black holes would introduce an additional, scale-dependent component to the total power spectrum \citep{Afshordi_2003} that could alter the abundance of collapsed halos, the star formation history, and the reionization process. In the following, we compute the total matter power spectrum as the sum of the $\Lambda$CDM and PBH contributions, normalized to reproduce the observed amplitude of matter fluctuations ($\sigma_8$) at $z = 0$, in order to quantify the imprint of primordial black holes on the growth of structure in the early Universe.
We follow the approach of \cite{Inman2019-fy,Matteri2025-wc} and include the PBH component to the power spectrum of $\Lambda$CDM as:
\begin{equation}
\label{eq:PS}
    P_{CDM} = P_{\Lambda CDM} + P_{PBH}
\end{equation}
where the additional term $P_{PBH}$ accounts for the Poissonian shot noise \citep{Meszaros75} produced by the discrete nature of PBHs.
In the case of an extended PBH mass function (\cite{Matteri2025-wc},  see Appendix \ref{sec:appendix_PS} for the derivation), the PBH power spectrum can be written as
\begin{align}
P_{PBH}(k) = \frac{D(z)^2}{\rho_c\; \Omega_{dm} } \int_{0}^{M(k)} {\frac{d\,f_{PBH}} {d\,\ln M}  dM},
\end{align}
where $D(z)$ is the growth factor as a function of redshift $z$, and $M(k)= 2 \pi^2 \Omega_{dm} {\rho_c}/{k^3}$ is the mass associated with a given critical wavenumber $k$.

\begin{figure}
    \centering
    \includegraphics[width=\linewidth]{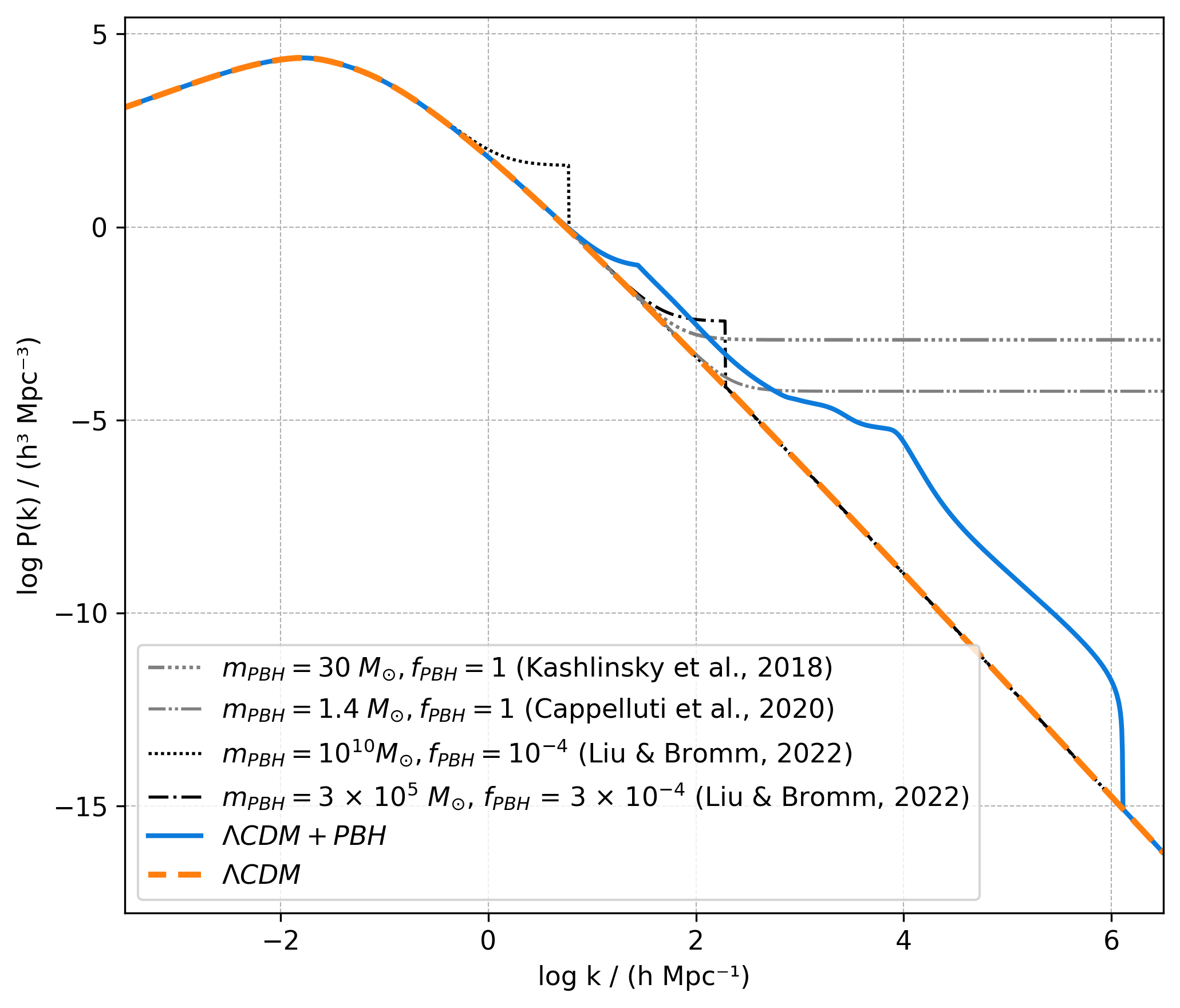}
    \caption{Matter power spectrum at $z = 0$ of $\Lambda$CDM + PBH component. The figure shows the baseline $\Lambda$CDM power spectrum with the dashed orange line, to which we added (separately) five different PBH power spectra: the two grey dash-dotted lines show two monochromatic PBH scenarios from \cite{Cappelluti2022,Kashlinsky2016}, and the two black dotted and dash-dotted curves correspond to two monochromatic PBH scenarios from \cite{Liu2022-xk} with a cutoff to the isocurvature term. The PBH power spectrum adopted in this paper and computed from the extended mass function in Figure \ref{fig:massfunction} is depicted with the solid, blue line.}
    \label{fig:PS}
\end{figure}

Starting from the PBH mass function shown in Figure~\ref{fig:massfunction}, we compute the corresponding PBH contribution to the matter power spectrum at $z=0$. The total power spectrum is then obtained by adding the $\Lambda$CDM and PBH components, after renormalizing their sum to match the observed value of $\sigma_8$ at $z=0$. The resulting spectrum is displayed in Figure~\ref{fig:PS}, where the blue solid curve denotes the $\Lambda$CDM+PBH case. 
For comparison, the orange dashed curve shows the baseline $\Lambda$CDM power spectrum, the two grey dash-dotted lines show two monochromatic PBH scenarios from \citep{Cappelluti2022,Kashlinsky2016}, and the black dotted and dash-dotted curves correspond to two monochromatic PBH scenarios from \citep{Liu2022-xk} where they impose a cutoff to the isocurvature term to suppress the power at scales smaller than $M \sim M_{PBH}$. We can see that the power spectrum as computed in this paper has non negligible, but yet important, contributions to the $\Lambda$CDM one. 
Compared to the monochromatic PBH cases mentioned above, which can produce sharp transitions and nearly scale-independent enhancements over part of the plotted range, the extended PBH mass function considered here gives a more gradual contribution to the total power spectrum. The resulting $\Lambda$CDM + PBH curve therefore remains smoother, with no abrupt plateau-like feature across the scales shown.

\subsection{Halo mass function}

The halo mass function describes the comoving number density $n(M_h,z)$ of dark matter halos as a function of their mass $M_h$ and redshift $z$, and it represents the key element of structure formation theory. It connects the statistical properties of the primordial density field to the nonlinear distribution of collapsed objects, thereby linking cosmological initial conditions to observable structures such as galaxies and clusters. In the standard cosmological framework, halo abundance is predicted using models such as the Press–Schechter formalism \citep{1974ApJ_PSformalism} and its extensions, which link the fraction of matter collapsed into halos to the variance of the linear density field smoothed on a given mass scale. 
However, achieving the accuracy required for precision cosmology demands empirical calibration from large N-body simulations. These simulations have shown that the mass function is not strictly universal: its amplitude and shape evolve systematically with redshift and depend on how halos are defined \citep{10.1046/j.1365-8711.1999.02692.x}. Numerical studies, such as those by \citep{Tinker2008-mh}, have refined the functional form of the mass function to percent-level precision across a wide mass and redshift range, revealing that the abundance of halos decreases monotonically with redshift and that the shape of the function evolves with the characteristic overdensity. This behavior reflects the combined influence of the cosmological growth rate, the evolution of halo concentrations, and the underlying matter power spectrum. 
We choose the following functional form to describe the DM halo mass function: 
\begin{align} 
\frac{dn}{dM_h}(M_h,z) = \Phi_{Tinker}(M_h,z) \times C_1(M_h,z) 
\end{align}
which is the product of the standard Tinker mass function \citep{Tinker2008-mh} times a correction factor \citep{Behroozi2013-do}. The full functional form of the halo mass function adopted in this work is reported in Appendix \ref{sec:appendix_halofunct}. We note that the presence of a PBH component modifies the halo mass function through its impact on the underlying matter distribution, as explicitly shown in Appendix \ref{sec:appendix_halofunct}.

\subsection{Star formation rate density}

\begin{figure}
    \centering
    \includegraphics[width=\linewidth]{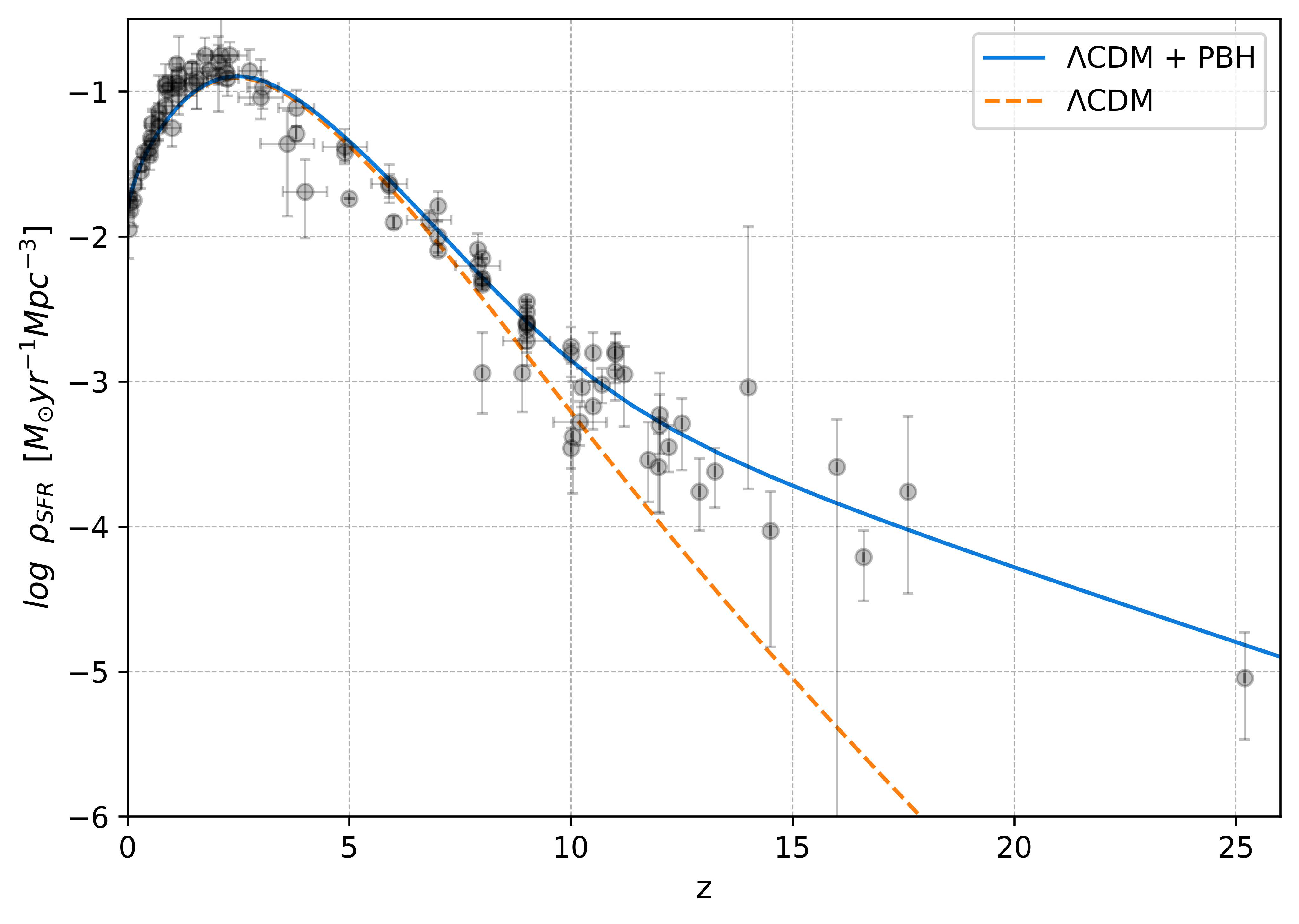}
    \caption{Star formation rate density as a function of redshift. The dashed orange curve shows the SFRD predicted by the standard $\Lambda$CDM matter power spectrum, while the solid blue curve corresponds to the SFRD obtained when including the contribution from primordial black holes. The data points represent observational SFRD measurements compiled from the literature \citep{Madau2014-rz,Oesch2014-us,Bouwens2016-ix,McLeod2016-ep,Oesch2018-sa,Donnan2022-jc,Finkelstein2023-dn,Harikane2023-ba,Bouwens2023-rl,McLeod2023-yh,Donnan2024-ry,Harikane2024-zn,Adams2024-md,Willott2024-cf,Perez-Gonzalez2025-sa}.}
    \label{fig:SFRD}
\end{figure}

The cosmic star formation rate (SFR) and its comoving integral, the star formation rate density (SFRD), quantify the rate at which baryons are converted into stars across cosmic time and therefore provide a direct record of galaxy assembly and feedback processes. The canonical picture compiled by \citep{Madau2014-rz} shows a steady rise of the SFRD from the earliest measurable epochs to a broad peak at redshift near two, followed by a decline to the present day; that compilation remains the baseline used to compare newer measurements. 
Early results from deep James Webb Space Telescope (JWST \footnote{\url{https://science.nasa.gov/mission/webb/}}) surveys have altered the observational landscape at the highest redshifts by revealing a population of UV-bright galaxies at $z \gtrsim 8-12$ whose abundance and luminosities in some analyses exceed simple extrapolations of the pre-JWST decline; these studies infer a less rapid falloff of the UV luminosity density and therefore a higher SFRD at $z \gtrsim 8$ than previously assumed (see the data points in Figure~\ref{fig:SFRD} and their corresponding references). 
The emerging JWST measurements have important implications for the timing and sources of reionization, the efficiency of star formation in low-mass halos, and the required feedback physics in galaxy formation models, but the quantitative SFRD at $z \gtrsim 8 $ is not yet settled and remains an active area of observational and theoretical work (e.g., \cite{Bouwens2023-rl}).

The star formation rate density ($\rho_{SFR}$ in the equations) can be expressed in the following way \citep{Harikane2022-rc}:
\begin{align}
\begin{split}
\label{eq:SFRD}
\rho_{\rm SFR}(z) &= \int \mathrm{SFR}(M_h,z)\,
\frac{dn}{d M_h}(M_h,z)\,d M_h =\\&= \int \frac{\mathrm{SFR}(M_h,z)}{\dot M_h}\,\dot M_h
\frac{dn}{d M_h}(M_h,z)\,d M_h,
\end{split}
\end{align}
where $\frac{dn}{d M_h}(M_h,z)$ is the modified Tinker halo mass function from the previous section, and $\mathrm{SFR}(M_h,z)/ \dot M_h$ represents the Star Formation Efficiency (SFE), which is the ratio between the SFR and the time derivative of the halo mass $M_h$. In our analysis, we choose the SFE as in \citep{Harikane2022-rc}. The computation of the SFRD follows the procedure outlined in Appendix \ref{sec:appendix_SFRD}, and the results are depicted in Figure \ref{fig:SFRD}, where the dashed orange line is the SFRD from the PS of $\Lambda$CDM alone, while the solid line is the PS with the PBH contribution (Eq. \ref{eq:PS}). Both lines are compared to experimental data. 
As evident from the figure, the inclusion of a PBH component leads to a systematic enhancement of the star formation rate density at higher redshift with respect to the standard $\Lambda$CDM prediction. While both models provide a comparable fit to the observational data at 
$z \lesssim 6$, the PBH scenario maintains a significantly higher SFRD at 
$z \gtrsim 8$, remaining closer to the currently available high-redshift measurements. This behavior reflects the additional small-scale power induced by PBHs, which promotes the earlier formation of low-mass halos and thus accelerates the onset of star formation. The overall agreement with current data, combined with the extended high-redshift tail, suggests that a PBH population can naturally alleviate tensions related to early structure formation without disrupting the standard cosmological framework at lower redshifts. 
We note, however, that an enhanced high-redshift SFRD may also arise if star formation was intrinsically more efficient at early times. In particular, feedback-free starbursts have been proposed as a mechanism to increase the star-formation efficiency in massive high-redshift halos, where the gas can reach sufficiently high densities that the free-fall time becomes shorter than the timescale for effective stellar feedback from winds and supernovae \citep{dekel+}. In this case, rapid gas collapse could allow a larger fraction of the accreted baryons to be converted into stars before feedback regulates the burst.
It is also important to note that part of the high-redshift JWST population may include compact accreting black holes rather than purely stellar systems. 
In particular, a fraction of the newly discovered broad-line AGN are classified as Little Red Dots (LRDs), objects interpreted as massive black holes embedded in dense, dust-poor or nearly pristine gas. The lensed LRD A2744-QSO1 at \(z=7.04\) illustrates how an early massive black hole can exist in an extremely chemically unevolved system \citep{10.1093/mnras/staf2109}.
Such sources may provide a possible observational counterpart of the high-mass tail of a PBH-seeded population, while remaining subject to significant modelling uncertainties.

\subsection{The Thomson optical depth}

The Thomson scattering optical depth, $\tau(z)$, measures the integrated probability that a CMB photon encounters free electrons between recombination and today; thus, it is an integral constraint on the total column of ionized gas produced during reionization. Current CMB polarization analyzes from Planck’s final 2018 release give $\tau = 0.054 \pm 0.007$ \citep{Planck_Collaboration2020-ij}, indicating late, relatively rapid reionization with most ionization occurring at $z \approx 7-9$; updated analyzes using the Planck PR4 maps find consistent values, with model-averaged constraints around $\tau = 0.058 \pm 0.006$ \citep{PlanckPR4}.
This integrated optical depth links directly to the time-dependent ionized volume fraction of hydrogen: the measured $\tau$ fixes the allowed integral of the free-electron history and therefore constrains how the ionized fraction 
must evolve.
The Thomson optical depth is defined as the integral of the electron density along the line of sight:
\begin{equation}
\tau(z) = c \, \sigma_T \int_{0}^{z} dz' \, \frac{n_e(z')}{(1+z')H(z')} ,
\end{equation}
where $c$ is the speed of light, $\sigma_T$ is the Thomson scattering cross section, $n_e(z)$ is the proper number density of free electrons, and $H(z) = H_0 \, \sqrt{\Omega_m (1+z)^3 + \Omega_\Lambda}$ is the expansion rate of the Universe. The exact procedure used to compute the Thompson optical depth adopted in this paper is outlined in Appendix \ref{sec:appendix_thompson}.

\begin{figure}
    \centering
    \includegraphics[width=\linewidth]{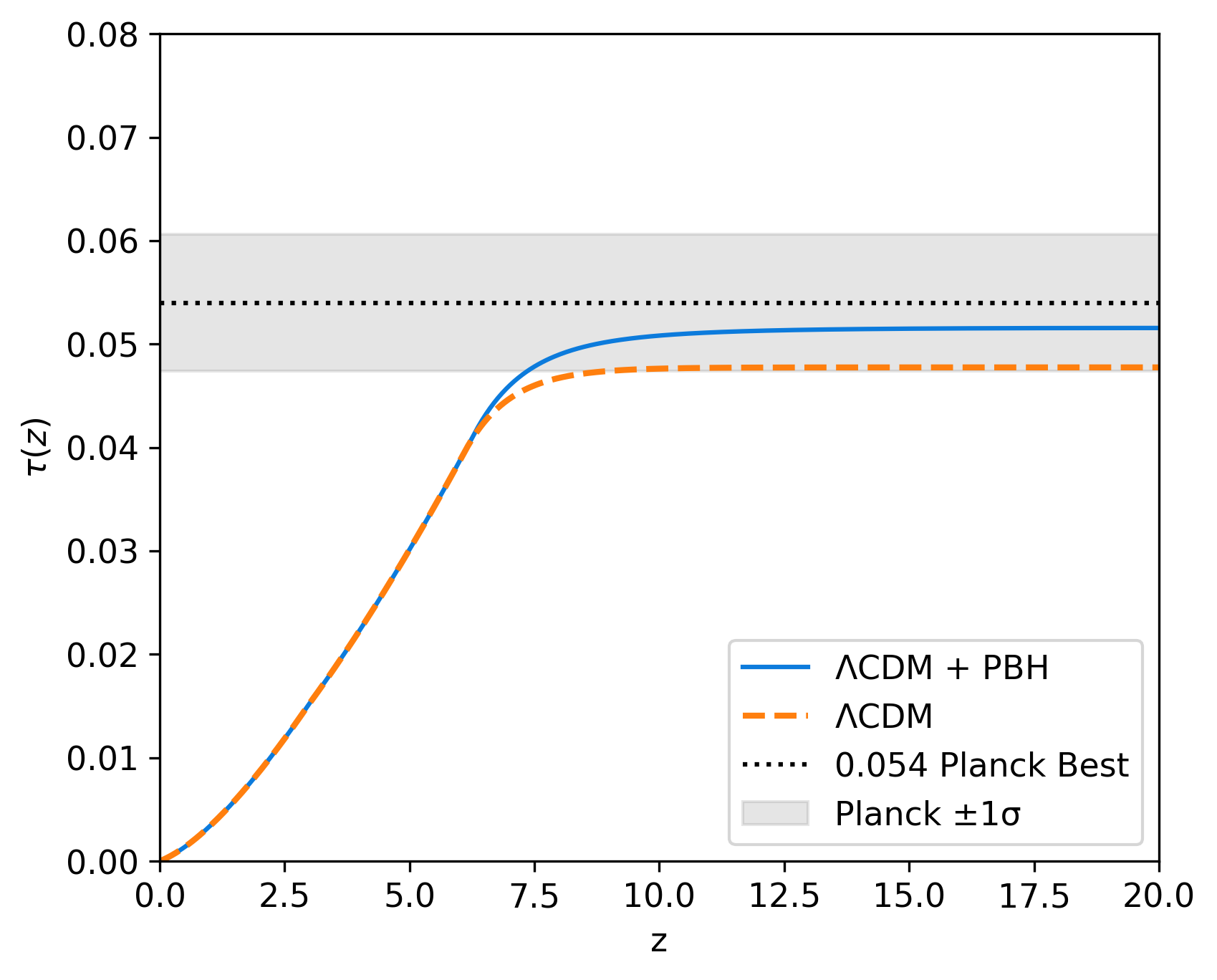}
    \caption{Thomson optical depth as a function of redshift. The dashed orange curve shows the prediction from the standard $\Lambda$CDM model, while the solid blue curve corresponds to $\Lambda$CDM including the contribution from primordial black holes. The horizontal dotted black line indicates the best-fit Thomson optical depth measured by Planck, with the shaded gray band denoting the $\pm1\sigma$ confidence interval.}
    \label{fig:thomson}
\end{figure}

The results for the Thomson optical depth as a function of redshift for the $\Lambda$CDM (dashed orange line) and $\Lambda$CDM+PBH (solid blue line) cases are shown in Figure~\ref{fig:thomson}. The inclusion of a PBH component leads to a modest increase in the Thomson optical depth at intermediate redshifts, reflecting the earlier onset of structure formation and the associated enhancement of ionizing sources. While the $\Lambda$CDM prediction remains systematically lower, both models converge at low redshift and exhibit a similar evolution at early times. Importantly, the PBH-enhanced scenario remains fully consistent with Planck constraints, with the predicted optical depth lying within the $\pm 1\sigma$ confidence interval and approaching the best-fit value. This behavior translates into an earlier and more extended reionization history, as shown in Figure~\ref{fig:reionizationhistory}, where the ionized fraction (as defined in Appendix \ref{sec:appendix_thompson}) is systematically higher at $z \gtrsim 6$ compared to the standard $\Lambda$CDM case. This is a direct consequence of the enhanced early star formation induced by the PBH component while remaining broadly consistent with current observational constraints.

\begin{figure}
    \centering
    \includegraphics[width=\linewidth]{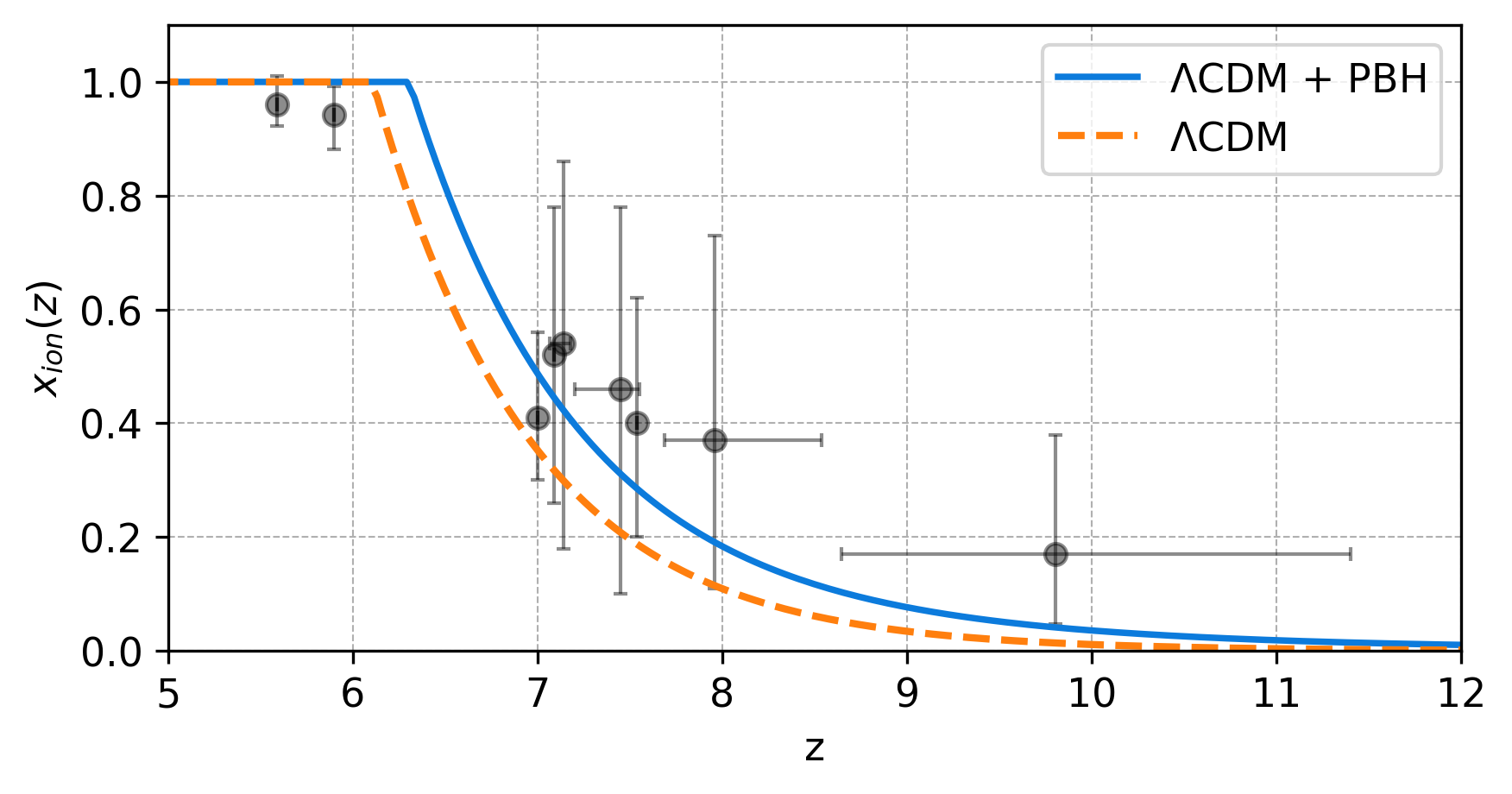}
    \caption{Reionization history expressed as the volume-averaged ionized hydrogen fraction, $x_{\rm ion}(z)$, as a function of redshift. The solid blue curve shows the prediction of the $\Lambda$CDM model including a PBH contribution, while the dashed orange curve corresponds to the standard $\Lambda$CDM scenario without PBHs. The inclusion of PBHs leads to an earlier and more extended reionization history. The black data points with error bars represent observational constraints compiled from the literature \citep{McGreer2014,Davies2018,Mason2018,Umeda2024}.}
    \label{fig:reionizationhistory}
\end{figure}

\subsection{PBH binaries}

Mergers of black holes represent one of the most powerful sources of gravitational radiation in the Universe. These events, involving the inspiral and the coalescence of compact objects, provide a unique probe of strong-field general relativity and the astrophysical processes that lead to black hole formation and evolution. The observed signals encode information about the component masses, spins, and orbital dynamics of the merging black holes. The rapidly expanding catalog of detections \citep{Abac2025} has revealed a population of stellar-mass black holes more diverse than previously thought, including objects with masses and spins that challenge predictions from standard stellar evolution theory \citep{Bird_2016}. 
In particular, several observed mergers involve black holes within or above the so-called “pair-instability mass gap” (around $50-120 \, M_\odot$), where stellar evolution models predict no remnants should form due to complete disruption of massive progenitors by pair-instability supernovae. Moreover, the apparent presence of systems with low effective spins suggests the existence of formation channels beyond conventional binary stellar evolution, such as dynamical assembly in dense stellar environments, hierarchical mergers, or the formation of PBH binaries, motivating new studies into their possible origins.
In this Section, we describe and characterize PBH mergers as a function of their masses as detectable by current and future interferometers.

PBH binaries can merge through two main formation channels, often referred to as the \emph{early} and \emph{late} populations.
The early merging channel originates from PBH pairs that decouple from the Hubble expansion in the radiation era, forming bound binaries prior to structure formation \citep{Raidal_2019}.
The late channel, on the other hand, describes PBH pairs which have a close encounter in the galaxy halo after structure formation and become a bound system with the emission of gravitational waves \citep{CLESSE2017105}. 
The two channels can in principle coexist, with each characterized by its own intrinsic merger rate density, $\tau_E(M_1,M_2)$ and $\tau_L(M_1,M_2)$, corresponding to the early- and late-formation mechanisms, respectively (see Subsubsections. \ref{subs:earlymergerrate} and \ref{subs:latemergerrate}).

\subsubsection{Early merger rate}

\label{subs:earlymergerrate}
Following \citet{Raidal_2019}, we describe the early--Universe PBH binary channel in terms of an intrinsic merger-rate density $\tau_E(m_1,m_2,z)\equiv {\rm d}^2R/{\rm d}m_1{\rm d}m_2$. In this picture, PBH binaries form during the radiation era when neighbouring PBHs gravitationally decouple from the Hubble flow; their initial orbital elements are set by tidal torques from the ambient PBH distribution and density perturbations, and the subsequent evolution to coalescence is assumed to be driven purely by gravitational--wave emission \citep{Raidal_2019}. For an extended PBH mass function we define:
$\psi(M) \equiv \frac{{\rm d}f_{\rm PBH}}{{\rm d}\ln M}$ and
$\langle M \rangle \equiv \int M\,\psi(M)\,{\rm d}\ln M $.
The merger-rate density can then be written as \citep{Raidal_2019}
\begin{equation}
\label{eq:Raidal}
\begin{split}
&\tau_E(M_1,M_2,z)=\\
&= \mathcal{A}\, f_{\rm PBH}^{\,\alpha}\,
\mathcal{F}(M_1,M_2,z)\,
S_1(M_{\rm tot})\,S_2(z)\,
\frac{\psi(M_1)\,\psi(M_2)}{\langle M\rangle^{2}},
\end{split}
\end{equation}
where $\mathcal{A}\simeq 1.6\times 10^{6}\ {\rm Gpc}^{-3}\,{\rm yr}^{-1}$, $\alpha=53/37$, and
\begin{equation}
\mathcal{F}(M_1,M_2,z)=
\left[\frac{t(z)}{t_0}\right]^{-34/37}
\left(\frac{M_{\rm tot}}{M_\odot}\right)^{-32/37}
\eta^{-34/37},
\end{equation}
with $M_{\rm tot}=M_1+M_2$ and $\eta=M_1M_2/M_{\rm tot}^2$.
When computing the intrinsic merger rate for a given binary, we approximate the abundance in a finite mass bin by
\(\psi(M_i) \simeq \left.\frac{{\rm d}f_{\rm PBH}}{{\rm d}\ln M}\right|_{M_i}\,\Delta\ln M\),
where $M_i$ is the representative mass of the bin and
\(\Delta\ln M = 0.1\,\ln 10\) is set by the grid spacing of the mass function. The factor $S_1$ describes the suppression due to nearby PBHs that can perturb or disrupt the primordial binary, while $S_2$ accounts for the late-time evolution of PBH clusters. In this framework, $S_1$ depends on the expected number of PBHs surrounding the binary and on the variance of density perturbations at matter--radiation equality, whereas $S_2$ captures the cumulative effects of PBH clustering and tidal interactions during the subsequent cosmic evolution. The two factors have rather complicated expressions, and can be found in Appendix \ref{sec:appendix_earlyMRD}.

The derivation of these factors assumes that primordial binaries form from nearest--neighbor PBHs during the radiation era, with their angular momentum generated by tidal torques from the surrounding PBH distribution and large-scale density perturbations. After formation, the binaries are assumed to evolve in isolation under gravitational-wave emission, while environmental interactions are treated statistically through the suppression factors described above.

\subsubsection{Late merger rate}
\label{subs:latemergerrate}

To characterize the late merger rate for an extended mass function, we follow the procedure outlined by \citep{Magaraggia2026} (their section 3.2), which is a generalization of the extended mass functions to the original computation by \citep{CLESSE2017105}. 

The cosmic mean dark-matter density is given by $\rho_{\rm DM}=\Omega_{\rm DM}\,\rho_c$. In galaxy haloes, where we assume the majority of PBHs reside, the local dark-matter density is enhanced by an overdensity factor $\delta_{\rm loc}$ relative to the cosmic mean, such that $\rho_{\rm DM}^{\rm loc}=\delta_{\rm loc}\,\rho_{\rm DM}$. Although the precise value of $\rho_{\rm DM}^{\rm loc}$ is uncertain, global analyses indicate that for Milky Way–like galaxies it lies in the range $\rho_{\rm DM}^{\rm loc}\simeq 0.3$–$0.5\;\mathrm{GeV\,cm^{-3}}$ \citep{deSalas_2021}. In the following, we adopt a fiducial value of $\rho_{\rm DM}^{\rm loc}=0.4\;\mathrm{GeV\,cm^{-3}}$, corresponding to $\delta_{\rm loc}\sim3\times10^{5}$.

We further assume that PBHs within haloes move with the characteristic dark-matter velocity of Milky Way–like systems. The velocity distribution is typically approximated as Maxwell–Boltzmann, with a peak at $v_{\rm vir}=250\;\mathrm{km\,s^{-1}}$ \citep{folsom2025darkmattervelocitydistributions}. Accordingly, we take the typical relative velocity between two PBHs to be $v=\sqrt{2}\,v_{\rm vir}$ \citep{PhysRevD.94.084013}.
These assumptions result in the following intrinsic late merger rate density for a given pair $(M_1,M_2)$:
\begin{equation}
    \tau_L(M_1,M_2)= A\; 
\frac{G^{2}\,\rho_{\rm DM}^{2}\,\delta_{\rm loc}^{2}\,f_{\rm PBH,1}f_{\rm PBH,2}}{c^{10/7}\,v^{11/7}}\frac{M_{tot}^{10/7}}{(M_1 M_2)^{5/7}}
\label{eq:tau_L}
\end{equation}
where $A=2\pi(\ln 10)^2\left(\frac{85\pi}{6\sqrt{2}}\right)^{2/7}$, and, similarly as before,
\(f_{\rm PBH,i} \simeq \left.\frac{{\rm d}f_{\rm PBH}}{{\rm d}\ln M}\right|_{M_i}\,\Delta\ln M\).

\subsubsection{GW mergers and the LVK detectable event rate}

The Laser Interferometer Gravitational-Wave Observatory, together with the Virgo and KAGRA interferometers, forms the LVK network, a global array of kilometer-scale laser interferometers designed to detect gravitational waves through the measurement of differential arm-length variations induced by passing spacetime perturbations. These detectors are most sensitive to compact binary coalescences with component masses ranging from a fraction of a solar mass to several tens of solar masses, corresponding to gravitational-wave frequencies from a few tens to a few thousand hertz \citep{Aasi_2015}. As a result, the LVK network primarily probes mergers of stellar-mass black holes and neutron stars, while retaining sensitivity to sub-solar mass binaries, which are of particular interest in searches for non-standard compact objects such as primordial black holes.
The detector response is characterized by the one--sided strain noise power spectral density $S_h(f)$, which is obtained from the amplitude spectral density ${\rm ASD}(f)$ through $S_h(f)=[{\rm ASD}(f)]^2$. In this work we adopt the O3b H1 amplitude spectral density of the LVK detector network\footnote{Publicly available at \url{https://dcc.ligo.org/LIGO-G2100672/public}}. 
The comoving volume effectively surveyed by an interferometer of this kind for a given binary of component masses $(M_1,M_2)$ is $V_{\rm det}(M_1,M_2)
= \frac{4\pi}{3}\,r_{\rm det}^3(M_1,M_2)$, 
where $r_{\rm det}(M_1,M_2)$ is the detector horizon distance, which is the luminosity distance of the farthest detectable source \citep{Carr2021-lr}.
The detected merger rate for a binary is therefore
\begin{equation}
\label{eq:R_det}
    R_{\rm LVK}(M_1,M_2)
    = \tau_L(M_1,M_2)\,V_{\rm det}(M_1,M_2),
\end{equation}
where $\tau_L(M_1,M_2)$ is the intrinsic late merger rate density. 
While in principle, both formation channels can contribute to the total merger rate density, in this subsection we focus only on the late PBH scenario, because it is expected to dominate at low redshift. The fraction of binaries that decouple in the early Universe is estimated to be of order $\mathcal{O}\,\left(0.01\, f_{\rm PBH}^{16/37}\right)$ \citep{PhysRevD.103.023026_wong}. Consequently, even in the limiting case $f_{\rm PBH}=1$, the formation of binaries in the early Universe is expected to remain a subdominant contribution in the low-redshift regime probed by LVK.

The maximum luminosity distance at which a binary can be detected, often referred to as the detector horizon distance $r_{\rm det}(M_1,M_2)$, can be written as \citep{Carr2021-lr}
\begin{align}
\begin{split}
    &r_{\rm det}(M_1,M_2)
    = \frac{\sqrt{5}\;c}{24\,\pi^{2/3}\,2.26}
      \left(\frac{G \mathcal{M}_c}{c^3}\right)^{5/6}
      \times \\
      &\times
      \left[
        \int_{f_{\min}}^{f_{\rm ISCO}(M_1,M_2)} 
          \frac{f^{-7/3}}{S_h(f)}\,{\rm d}f
        + \int_{f_{\rm ISCO}(M_1,M_2)}^{f_{\max}(M_1,M_2)}
          \frac{f^{-2/3}}{S_h(f)}\,{\rm d}f
      \right]^{1/2}
    \label{eq:detector}
\end{split}
\end{align}
where $\mathcal{M}_c = \frac{(M_1 M_2)^{3/5}}{(M_1+M_2)^{1/5}}$ is the chirp mass and the factor $2.26$ accounts for the averaging over source orientations in an Euclidean Universe.

In principle, the waveform should include cosmological redshift corrections by replacing $\mathcal{M}_c$ with $\mathcal{M}_c(1+z)$. However, for the low--mass binaries considered here, the detectable volume lies predominantly in the nearby Universe ($z\lesssim0.05$), so redshift effects introduce only percent-level corrections, which are subdominant compared to the astrophysical uncertainties of the model \citep{Carr2021-lr}. 

The terms inside the square brackets describe the contributions from different phases of the gravitational-wave signal across the detector bandwidth. The first integral accounts for the inspiral phase, whose post--Newtonian waveform scales as $f^{-7/3}$ between the minimum frequency detectable by the interferometers ($f_{\min}=10\,\mathrm{Hz}$ for LVK O3b) and the innermost stable circular orbit frequency $f_{\rm ISCO}=\frac{4400}{M_{\rm tot}}\ \mathrm{Hz}$, where $M_{\rm tot}$ is expressed in solar masses. The second term describes the merger phase, with frequency dependence $f^{-2/3}$ between $f_{\rm ISCO}$ and
$f_{\max}=\min\left(f_{\rm merge},\,2000\ \mathrm{Hz}\right)$,
where the merger frequency is given by
$\pi M_{\rm tot} f_{\rm merge} = a_0 \eta^{2} + b_0 \eta + c_0$ \citep{Ajith_2008},
with $\eta$ the symmetric mass ratio
and coefficients
$a_0 = 6.6389\times10^{-1}, 
b_0 = -1.0321\times10^{-1},
c_0 = 1.0979\times10^{-1}$.

\begin{figure}
    \centering
    \includegraphics[width=\linewidth]{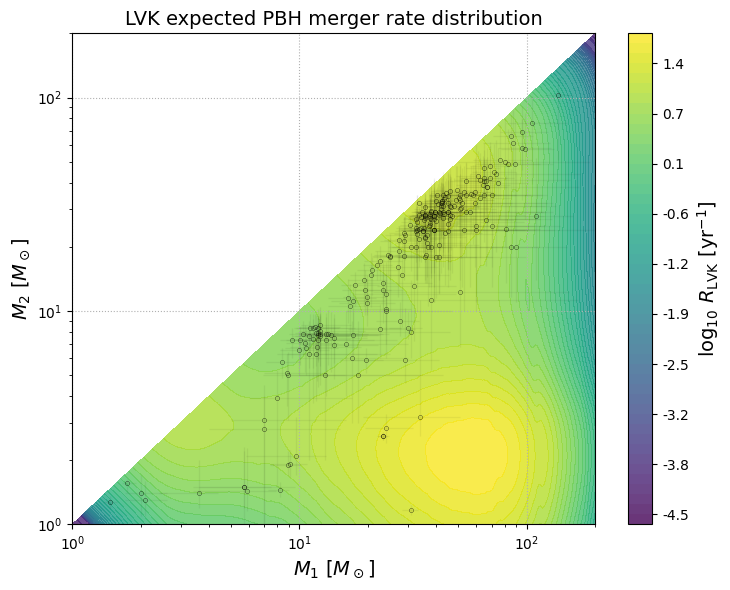}
    \caption{Distribution of the predicted PBH binary merger detection rate for the LVK network in the $(M_1,M_2)$ plane, where $M_1 \geq M_2$. The color scale shows the expected detection rate $\log_{10} R_{\rm LVK}$ (yr$^{-1}$) for binaries with component masses $(M_1,M_2)$, computed under the assumed PBH mass function and detector sensitivity. The triangular region corresponds to the physically allowed parameter space $M_1 \geq M_2$. Circles with gray error bars (publicly available at \url{https://gwosc.org/eventapi/html/allevents/}) indicate the component masses of binary black hole mergers reported by the LVK collaboration, with their associated uncertainties.}
    
    \label{fig:LVK}
\end{figure}

We now have all the ingredients to compute the expected rate as seen from LVK as a function of the two BH masses. The results can be seen in Figure \ref{fig:LVK}, which shows the predicted LVK detection rate for primordial black hole (PBH) binary mergers in the plane of the component masses $(M_1,M_2)$, assuming $M_1 \geq M_2$. The color scale represents the logarithm of the expected detection rate, $\log_{10} R_{\rm LVK}$ (yr$^{-1}$), computed for binaries with masses $(M_1,M_2)$ under the PBH population model adopted in this work. The triangular region corresponds to the physically allowed parameter space where the primary mass exceeds the secondary mass. Superimposed are the binary black hole mergers detected by the LIGO–Virgo–KAGRA collaboration, shown as circles with their corresponding mass uncertainties.
The predicted detection-rate map exhibits a broad maximum at primary masses of a few tens of solar masses and secondary masses of a few to $\sim 30\,M_\odot$. The PBH population model considered here can naturally produce mergers in the mass range where LVK has seen most events. In this sense, the distribution of detected events does not contradict the expectations of the model and is consistent with the idea that PBH binaries could contribute to the observed merger population. However, some tensions are also visible. The predicted rate map extends toward lower secondary masses and lower mass ratios where relatively few LVK detections are currently observed. 
In the late-capture channel adopted here, the intrinsic merger rate contains an explicit mass-dependent factor,
\(\tau_L\propto
M_{\rm tot}^{10/7}/(M_1M_2)^{5/7}\). For fixed primary mass and mass ratio $q$, this term scales as \((1+q)^{10/7}q^{-5/7}\), and therefore intrinsically enhances unequal-mass encounters. In addition, the predicted distribution is sensitive to the relative normalization of the PBH mass function in the Chandrasekhar/sub-solar regime and in the tens-of-solar-mass regime: even if the lepton-flavour asymmetry suppresses the original \(\sim 1.5\,M_\odot\) peak, any residual abundance in this mass range can efficiently pair with the higher-mass PBH component and populate the low-\(q\) region. Detector selection partially compensates this effect, since unequal-mass systems have smaller chirp masses and therefore smaller detectable volumes, but it does not necessarily remove the intrinsic low-\(q\) enhancement. The apparent mismatch with the LVK population may therefore indicate that the low-mass part of the adopted PBH mass function, the late-capture prescription, or both, overproduce asymmetric binaries.
Only future data can settle this issue.

\subsubsection{LISA detectable event rate}

The Laser Interferometer Space Antenna is a planned space-based gravitational-wave observatory that will detect gravitational waves by monitoring changes in the separations between three spacecraft arranged in a triangular constellation 2.5 million kilometers apart, using laser interferometry \citep{amaroseoane2017laserinterferometerspaceantenna}. Operating in the milli-hertz frequency band, LISA will be sensitive to sources that are inaccessible to ground-based detectors, including mergers of massive and supermassive black holes, extreme mass-ratio inspirals, and compact binaries with long orbital periods in the Milky Way \citep{amaroseoane2017laserinterferometerspaceantenna}. By probing these lower frequencies, LISA will provide a complementary view of gravitational-wave sources and their evolution across cosmic time, linking the growth of black holes to galaxy formation and large-scale structure. In this section, we focus on more massive black holes (MBHs) that span the mass range of
$\sim 10^{3}$--$10^{8}\,M_\odot$, because this is where LISA will be most sensitive and where the different formation pathways for supermassive black holes lead to distinctive predictions for the mass--redshift distribution of mergers \citep[see for instance][]{Ricarte+2018}. 
In astrophysical scenarios, \textit{light seeds} originate from Population~III 
stellar remnants with $M_{BH} \sim 10^{2}\,M_\odot$ \citep{Volonteri2010} and can grow into the LISA band only after sustained accretion and hierarchical mergers, 
implying that detectable MBH binaries are expected predominantly at $z\lesssim 15$ and with masses typically below $\sim 10^{6}\,M_\odot$. 
In contrast, \textit{heavy seeds} form through direct-collapse episodes with initial masses $M_{BH} \sim 10^{4}$--$10^{6}\,M_\odot$ \citep{Volonteri2010,Ricarte+2018}, allowing 
LISA-band mergers to occur earlier, at $z\sim 10$--$20$, and producing 
a characteristic concentration of events in the $10^{4}$--$10^{6}\,M_\odot$ mass range.  Primordial-black-hole scenarios differ qualitatively: the PBH mass function can populate the $\gtrsim 10^{4}\,M_\odot$ regime from the start, without relying on baryonic processes or halo assembly. As a result, PBH models naturally predict a non-negligible merger rate for very massive BH binaries even at $z\gtrsim 20$--$30$, extending into mass ranges and epochs that are inaccessible to astrophysical seeds. This leads to a clean theoretical separation in the $(M,z)$ plane: 
(i) at $z<10$ all channels may contribute; 
(ii) in the interval $10<z<20$ mergers of $10^{4}$--$10^{6}\,M_\odot$ black holes are characteristic of heavy seeds, whereas PBHs may also contribute if their high-mass tail is sufficiently populated; 
(iii) at $z>20$ astrophysical channels may be strongly suppressed by baryonic cooling, halo-assembly times, and Eddington-limited growth, so any LISA detection of MBH mergers in this regime (especially at masses $\gtrsim 10^{4}\,M_\odot$) would strongly favor a primordial origin. We note recent work on the Not-Quite Primoridal Black Holes (NQPBHs) by \citep{Qin+2025}, an alternative seeding scenario, demonstrating that direct collapse can lead to formation of BH seeds at $z \sim 100$ that are also expected to produce LISA events at these epochs. The combined mass and redshift distribution that LISA will measure stands to provide a powerful diagnostic of SMBH formation mechanisms.

From the detector point of view, LISA differs fundamentally from ground-based observatories such as LIGO-Virgo-KAGRA. LVK mainly observes stellar-mass binaries within a few Gpc, so that a Euclidean approximation for the detectable volume is often adequate.
LISA meanwhile can detect MBH mergers up to $z \sim 30$ and even beyond, depending on the mass range selected. In this regime, a naive Euclidean scaling is no longer valid: (i) the luminosity distance $D_L(z)$ must be computed in a specific cosmology, (ii) the detector sees redshifted, not rest-frame, masses ($\mathcal{M}_z = (1+z)\mathcal{M}_c$), and (iii) the observed event rate is diluted by time dilation. For these reasons, we cannot simply transplant the last section's Euclidean $V_{\rm det}$ approach to LISA, but we must compute the signal-to-noise ratio and the detectable volume in a fully cosmological way, which we do below. 

The detector selection is imposed by computing, for each $(M_1,M_2,z)$,
the sky-averaged signal-to-noise ratio expected in LISA.
We adopt a frequency-domain approximation for the GW waveform and define the redshifted chirp mass
$\mathcal{M}_z = (1+z)\,\mathcal{M}_c$.
Using the LISA strain noise power spectral density $S_n(f)$,
the signal-to-noise ratio is evaluated as
\begin{align}
\begin{split}
&\rho^2(M_1,M_2,z)
=
\frac{2}{3}\,\pi^{-4/3}
\frac{(G \,\mathcal{M}_z)^{5/3}}{c^3 \,D_L(z)^2} \times
\\ &\times \Bigg[
\int_{f_{\rm min}}^{f_{\rm ISCO}/(1+z)}
\frac{f^{-7/3}}{S_n(f)}\,{\rm d}f
+
\int_{f_{\rm ISCO}/(1+z)}^{f_{\rm max}}
\frac{f^{-2/3}}{S_n(f)}\,{\rm d}f
\Bigg],
\end{split}
\end{align}
where $D_L(z)= (1+z)\,c\int_0^z \frac{{\rm d}z'}{H(z')}$ is the luminosity
distance in the adopted cosmology.

The integration limits depend on the LISA frequency band and on the
binary evolution. The upper limit is defined as
$f_{\rm max} = \min\!\left(\frac{f_{\rm merge}}{1+z},\, f_{\max}^{\rm LISA}\right)$,
i.e. the minimum between the redshifted merge frequency (the same expression as in the LVK section) and the
high-frequency edge of the LISA band.
The lower limit is instead
$f_{\rm min} = \max\!\left(f_{\min}^{\rm LISA},\, f_{\rm start}\right)$,
where we adopt $f_{\min}^{\rm LISA}=10^{-4}\,{\rm Hz}$ and
$f_{\max}^{\rm LISA}=1\,{\rm Hz}$ for the LISA bandpass.
The start frequency $f_{\rm start}$ corresponds to a look-back time equal to the effective mission duration $T_{\rm obs}=4$ years and is obtained from the leading-order inspiral relation
$T_{\rm obs} =
\frac{5}{256}\,
\mathcal{M}_z^{-5/3}
(\pi f_{\rm start})^{-8/3}$.
As seen for LVK before, the two terms in the square brackets describe the contributions from
different stages of the waveform: the first integral corresponds to the inspiral phase with the characteristic frequency dependence
$f^{-7/3}$, while the second term accounts for the transition toward the merger regime, approximated here with a scaling $\propto f^{-2/3}$. Only binaries satisfying the detection condition $\rho(M_1,M_2,z) \ge 9$ are included in the detected population.

In our framework, the intrinsic merger-rate density is determined by the early-Universe PBH binary formation channel described in Subsect. \ref{subs:earlymergerrate}. We note that we evaluate the merger rate only for the high-mass portion of the PBH mass function, restricting the computation to $M \gtrsim 10^{3}\,M_\odot$. This choice is motivated by the fact that our analysis focuses on the gravitational-wave signal detectable by LISA, whose sensitivity peaks in the intermediate- and supermassive black-hole regime. Lower-mass PBHs present in the full mass spectrum, therefore, contribute negligibly to the LISA detection rate considered here. Within this restricted mass range the PBH mass fraction is small compared to the total PBH abundance, which implies that the typical number of neighbouring PBHs capable of perturbing a primordial binary is reduced.

Since LISA is expected to probe mergers at high redshift, when large-scale structure is not yet fully established, the late formation channel associated with structure formation cannot be applied. Consequently, in this regime we rely exclusively on the early-Universe channel to produce binaries that can merge at such early cosmic times. The observable merger rate is then obtained by integrating the  early intrinsic rate over the cosmological volume accessible to the detector and applying the SNR selection.
For given redshift and mass intervals, the total LISA detection rate is then given by:
\begin{equation}
\begin{split}
&R_{\rm LISA}
=
\int_{z_{\min}}^{z_{\max}}
\frac{{\rm d}V_c}{{\rm d}z}
\frac{{\rm d}z}{1+z} 
\int_{M_{\min}}^{M_{\max}}
\int_{M_{\min}}^{M_{\max}}\\
&\tau_E(M_1,M_2,z)\,
\Theta\!\left[\rho(M_1,M_2,z)-\rho_{\rm thr}\right]
\,{\rm d}M_1\,{\rm d}M_2 ,
\end{split}
\end{equation}
where $\frac{{\rm d}V_c}{{\rm d}z}=
4\pi\,D_c^2(z)\,
\frac{c}{H(z)}$ is the differential comoving volume
element and the factor $(1+z)^{-1}$ accounts for the conversion
between source-frame and observer-frame time. This procedure
replaces the Euclidean detectable-volume approximation by a fully cosmological treatment that is appropriate for LISA’s high-redshift.

\begin{figure}
    \centering
    \includegraphics[width=\linewidth]{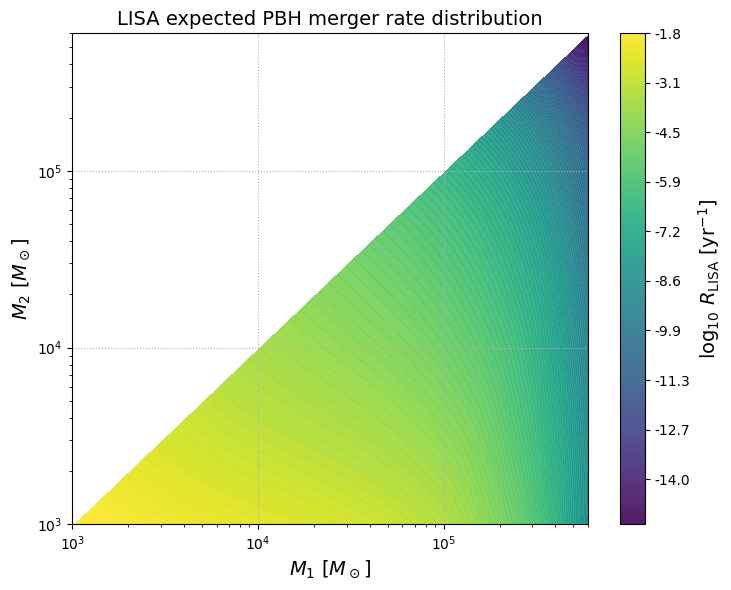}
    \caption{Predicted distribution of PBH binary mergers detectable by LISA in the $(M_1,M_2)$ plane, assuming $M_1 \geq M_2$. The color scale shows the logarithm of the cumulative detection rate, $\log_{10} R_{\rm LISA}$ (yr$^{-1}$), integrated over the redshift interval $5 \leq z \leq 20$. The triangular region corresponds to the physically allowed parameter space where the primary mass exceeds the secondary mass. The distribution reflects the combined effect of the assumed PBH mass function, the intrinsic merger rate, and the LISA selection function. The highest predicted detection rates occur for binaries with masses of order $10^3$–$10^4\,M_\odot$, while the rate decreases toward higher masses.}
    \label{fig:LISA}
\end{figure}

Figure \ref{fig:LISA} shows the predicted distribution of PBH binary mergers detectable by LISA in the plane of the component masses $(M_1,M_2)$, with $M_1 \geq M_2$. The color scale represents the logarithm of the cumulative detection rate, $\log_{10} R_{\rm LISA}$ (yr$^{-1}$), integrated over the redshift interval $5 \leq z \leq 20$. Each point in the diagram therefore indicates the expected number of LISA detections per year arising from PBH binaries with component masses $(M_1,M_2)$ within this redshift range, after accounting for the assumed PBH mass function and the LISA sensitivity to massive black hole inspirals.

The predicted distribution spans binaries with masses between $\sim10^3$ and a few $10^5\,M_\odot$, corresponding to the mass range where LISA is most sensitive to inspiral signals. The detection rate is highest toward the lower-left region of the diagram, where both component masses are relatively low within the LISA band (i.e., $M_1 \sim 10^3$–$10^4\,M_\odot$ and comparable $M_2$). In this region, each individual \((M_1,M_2)\) grid cell can contribute at the level of \(10^{-2}\)--\(10^{-3}\,{\rm yr}^{-1}\), reflecting the combined effect
of the PBH mass function, the intrinsic merger rate, and the larger number
density of lower-mass systems.

Moving toward higher masses, the predicted rate decreases progressively, reaching values well below $10^{-6}$ yr$^{-1}$ for binaries approaching $\sim10^5$–$10^6\,M_\odot$. This decline arises from two effects. First, the adopted PBH mass function predicts a rapidly decreasing abundance toward these higher masses. Second, very massive binaries spend less time in the LISA frequency band during the inspiral phase, reducing the probability that they will be observed during the mission lifetime.

Overall, the model predicts that if PBHs exist in this mass range, LISA detections should be dominated by mergers involving intermediate-mass black holes with masses of a few $10^3$–$10^4\,M_\odot$, with the event rate declining steadily toward higher masses and more extreme mass ratios.

Table \ref{table:detectedmergers} provides a more quantitative view of the trends illustrated in Figure \ref{fig:LISA} by reporting the cumulative merger rates detectable by LISA after summing over all \((M_1,M_2)\) grid cells contained in
each mass and redshift interval. The parameter space is divided into three mass bins, expressed in terms of $\log_{10}(M/M_\odot)$, and three redshift ranges spanning $5<z<50$.

The results show that the predicted LISA detections are strongly dominated by binaries with total masses in the range $10^3 \lesssim M/M_\odot \lesssim 10^4$. In this mass bin, the model predicts of the order a few to $\sim 10$ detectable mergers per year in the redshift interval $5<z<20$, with the rate gradually decreasing toward higher redshifts, reaching a few events per year at $35<z<50$. This behavior reflects both the intrinsic merger-rate evolution and the reduced detectability of more distant sources.

In contrast, the predicted detection rate drops sharply for higher masses. In the interval $10^4 \lesssim M/M_\odot \lesssim 10^5$, the expected rate falls below $\sim 0.1$ yr$^{-1}$ across all redshift bins. Finally, for systems with $M \gtrsim 10^5\,M_\odot$, the predicted detection rate becomes effectively negligible in our model.

Overall, these results indicate that within the PBH scenario considered here, LISA is expected to detect only a limited number of mergers involving intermediate-mass primordial black holes. The detectable population is predicted to be dominated by systems with masses of a few $10^3$–$10^4\,M_\odot$, while mergers involving more massive PBHs are expected to be extremely rare. This quantitative breakdown therefore reinforces the qualitative trends shown in Figure \ref{fig:LISA} and highlights that, in our framework, LISA would probe only a small population of intermediate-mass PBH mergers.

We emphasize, however, that this conclusion applies specifically to the early PBH binary formation channel considered here. In particular, the absence of detectable mergers at $M \gtrsim 10^5\,M_\odot$ does not imply that supermassive black hole binaries would not merge in general at lower redshifts. Rather, it indicates that mergers of such massive systems are not expected to originate from the primordial binary population formed in the early Universe. Supermassive black hole binaries produced through standard astrophysical processes, such as galaxy mergers and subsequent dynamical evolution, are still expected to occur and to contribute significantly to the merger population observed by LISA, but later in the Universe.
Published predictions for LISA MBH-binary detections span a wide model-dependent range, from a few to several hundred events over a multi-year mission \citep{Sesana_2009,PhysRevD.111.063051_Sadiq}. As a recent estimate, \cite{10.105100046361202556833}, using a light-seed semi-analytic model tied to hierarchical galaxy assembly, find LISA detection rates of \(0.29\)--\(14.91\,{\rm yr}^{-1}\) for \(M_{\rm bin}\in[10^3,10^5]\,M_\odot\) and \(0.16\)--\(11.08\,{\rm yr}^{-1}\) for \(M_{\rm bin}\in[10^5,10^7]\,M_\odot\), depending on their pessimistic or optimistic treatment of dynamical-friction delays.
Although these rates can be comparable to the PBH rates predicted here in the \(10^3\)--\(10^4\,M_\odot\) range, the redshift distribution is qualitatively different: astrophysical light-seed mergers are expected mainly at lower redshift, whereas the PBH channel considered here can yield detectable mergers at \(z\gtrsim 20\) and up to \(z\sim 50\). The redshift distribution is therefore the key diagnostic: the detection of black hole mergers at very high redshift, regardless of their masses, would be difficult to explain within standard astrophysical formation scenarios. Such events would therefore provide strong evidence for the existence of PBHs, and could indicate that the earliest supermassive black holes originate from the very massive tail of the PBH mass function.

\begin{table}[t]
\label{table:detectedmergers}
\centering
\caption{LISA detectable merger rates (yr$^{-1}$) predicted by our PBH model in the relevant mass and redshift intervals.}
\vspace{0.3cm}
\renewcommand{\arraystretch}{1.35}
\begin{tabular}{c|ccc}
\hline\hline
$\log_{10}(M/M_\odot)$ 
    & $5<z<20$ & $20<z<35$ & $35<z<50$ \\
\hline
$3 \le \log_{10} M < 4$ 
    & $9$ 
    & $5$
    & $3$ \\
$4 \le \log_{10} M < 5$
    & $0.07$
    & $0.04$
    & $0.03$ \\
$ \log_{10} M > 5$
    & 0
    & 0
    & 0\\
\hline\hline
\end{tabular}
\end{table}

\section{Discussion}

The results presented here demonstrate that an extended PBH mass function, motivated by features in the thermal history and equation of state of the early Universe, can leave correlated imprints across cosmological structure formation and gravitational-wave observables. Starting from a physically motivated PBH spectrum, we have propagated its consequences into the matter power spectrum, halo mass function, cosmic star-formation-rate density, reionization history, and the expected binary merger rates for current and future gravitational-wave detectors. The central question is whether a single PBH population can remain consistent with existing cosmological constraints while producing detectable signatures in gravitational-wave experiments. Our analysis shows that this is possible, but in a constrained and highly diagnostic way: the effects on structure formation are moderate, while the gravitational-wave signatures are most promising in specific regions of mass and redshift space.

\subsection{Impact on structure formation}

The PBH population modifies the matter power spectrum through the additional Poisson contribution generated by the discrete nature of compact dark-matter components. For an extended PBH mass function, this contribution is spread over a broad range of scales, yielding a smooth enhancement of small-scale power rather than the sharper scale-dependent features associated with monochromatic PBH models. Consequently, the total matter power spectrum remains close to the standard $\Lambda$CDM prediction on large scales while acquiring a non-negligible excess at small scales.

This excess power increases the abundance of low-mass halos at early times. The effect is, however, controlled in amplitude: it does not qualitatively alter the hierarchical structure-formation picture, nor does it erase the successful large-scale predictions of $\Lambda$CDM. Instead, the PBH component acts as a subdominant accelerator of early collapse, preferentially enhancing the earliest low-mass halos while leaving the global halo population broadly consistent with the standard cosmological framework. This is an important consistency check since any viable PBH scenario must generate observable signatures without overproducing small-scale structure.

\subsection{Implications for early star formation and reionization}

The enhancement in the early low-mass halo abundance propagates naturally into the baryonic sector. In our model, the PBH contribution produces a modest increase in the high-redshift star-formation rate density relative to the prediction obtained from the $\Lambda$CDM power spectrum alone. This increase is most relevant at the earliest epochs, where small changes in halo abundance can translate into an appreciable change in the onset of star formation. The resulting star-formation history remains broadly compatible with current observational estimates, including recent high-redshift measurements that suggest a slower decline in the abundance of early galaxies.

The corresponding effect on reionization is similarly moderate. The PBH-enhanced model produces a somewhat earlier and more extended ionization history, but the predicted Thomson optical depth remains consistent with current Planck constraints. This is a critical point: the PBH-induced increase in early star formation does not overproduce ionizing photons. The scenario, therefore, occupies an interesting regime in which PBHs can mildly advance the timing of early structure formation and reionization while remaining observationally viable.

\subsection{Interpretation of stellar-mass mergers observed by LVK}

Ground-based gravitational-wave observations probe the stellar-mass portion of the PBH mass function. In the LVK-accessible regime, the predicted detection-rate distribution overlaps substantially with the region populated by observed binary black-hole mergers. The adopted PBH mass function therefore naturally produces binaries in the mass range where the LVK network is most sensitive and where many detections are currently located.

This agreement should be interpreted carefully. The observed LVK population is not, by itself, evidence for a primordial origin, since stellar evolution and dynamical assembly channels can also populate this mass range. Rather, our results show that the PBH scenario is not excluded by the current mass distribution and could contribute to the observed compact-binary population. At the same time, the analytic PBH merger-rate model tends to favor systems with relatively unequal component masses, whereas many LVK binaries are closer to equal mass. This difference may reflect detector selection effects, uncertainties in the PBH mass function, or the incompleteness of the simplified merger prescription adopted here. It may also indicate that the LVK catalog is heterogeneous, containing both astrophysical and primordial binaries. Thus, PBHs are best interpreted as a possible contributing subpopulation rather than as an explanation for all LVK events.

\subsection{Predictions for massive black-hole mergers detectable by LISA}

LISA will probe a complementary regime, opening access to intermediate- and massive-black-hole mergers at high redshift. This mass--redshift window is especially powerful for testing PBH scenarios because primordial black holes can occupy the intermediate-mass regime from the earliest cosmic epochs, without requiring stellar evolution, baryonic cooling, or sustained accretion.

Our calculations show that the LISA-detectable PBH merger population is dominated by binaries with component masses of order $10^3$--$10^4\,M_\odot$. These systems lie near the optimal sensitivity range of LISA and correspond to the most populated portion of the high-mass tail of the adopted PBH mass function. The predicted detection rate decreases rapidly at higher masses, both because the PBH abundance falls and because more massive systems spend less time in the LISA band. Quantitatively, the expected rate is of order a few events per year for $10^3 \lesssim M/M_\odot \lesssim 10^4$ in the redshift interval $5<z<20$, with lower rates at higher redshift and negligible rates for substantially larger masses.

These predictions apply specifically to binaries formed through the early-Universe PBH channel considered here. The lack of detectable primordial binaries above $\sim 10^5\,M_\odot$ should not be interpreted as the absence of massive black-hole mergers in general. Astrophysical processes associated with galaxy assembly will still generate massive and supermassive black-hole binaries at later times. Rather, the result implies that, within this PBH framework, LISA is expected to detect only a limited but potentially distinctive population of intermediate-mass primordial black-hole mergers.

\subsection{Implications for the origin of massive black holes}

The origin of the first massive black holes remains a central problem in astrophysics. Stellar-remnant seeds, direct-collapse black holes, dense stellar systems, and primordial black holes each predict different mass functions, redshift distributions, and merger histories. Gravitational-wave observations provide a uniquely direct way to discriminate among these channels.

In this context, the most decisive signature would be the detection of intermediate-mass black-hole mergers at very high redshift. Such events would be difficult to accommodate in purely astrophysical scenarios, where the formation and growth of massive seeds require halo assembly, baryonic cooling, accretion, and dynamical pairing. PBHs, by contrast, can populate the intermediate-mass regime ab initio. A LISA detection of $10^3$--$10^4\,M_\odot$ mergers at $z \gtrsim 20$, and especially at redshifts approaching $z \sim 50$, would therefore constitute a strong indication of a primordial origin.

More broadly, the combined cosmological and gravitational-wave signatures provide a multi-scale test of the PBH hypothesis. The same PBH population that mildly enhances early small-scale structure also predicts a specific gravitational-wave phenomenology: a possible contribution to the LVK stellar-mass population and a rare, high-redshift population of intermediate-mass mergers detectable by LISA. This complementarity is the main strength of the framework. PBHs are not merely dark-matter candidates in this picture; they are fossil tracers of early-Universe physics, whose mass spectrum, clustering, and merger history encode conditions inaccessible to electromagnetic observations alone.

Several uncertainties remain. The adopted extended mass function is phenomenological and depends on the detailed thermodynamic history of the early Universe, including the treatment of particle thresholds, phase transitions, and possible lepton asymmetries. The mapping between extended PBH mass functions and existing monochromatic constraints also remains nontrivial, requiring observable-specific convolution over the full mass distribution. On the gravitational-wave side, the predicted rates depend on simplified prescriptions for early and late PBH binary formation, environmental disruption, clustering, and detector selection. Future work should therefore couple more complete PBH formation calculations to improved merger-rate modeling and a fully consistent treatment of observational constraints.

Despite these caveats, the main conclusion is robust: an extended PBH mass function can produce modest, observationally allowed changes to early structure formation while yielding distinctive gravitational-wave signatures. The most compelling tests will come from combining high-redshift galaxy and reionization measurements with the evolving LVK catalog and, ultimately, LISA detections. In particular, even a small number of high-redshift intermediate-mass black-hole mergers would provide a qualitatively new probe of primordial black holes and of the physical conditions that shaped the earliest phases of cosmic structure formation.

\section*{Data Availability}
The Jupyter Notebook and the data points to reproduce the figures in this paper are publicly available at \href{https://github.com/albertomagaraggia/PBH_CosmicArchitects}{Github}. To ensure full reproducibility, each subsection of this paper corresponds to a Jupyter Notebook section with the same title, allowing direct correspondence between the presented analysis and its computational implementation.
\section*{Acknowledgments}
A.M. acknowledges insightful discussions with Alberto Salvarese, Aidan J. Kaminsky, and Russell F. Roberts. 
A.M. thanks Deutsches Zentrum f\"ur Astrophysik and the Interdisciplinary Centre for Transformative Urban Regeneration for kind hospitality in Summer 2025. 
M.G. thanks Julien Froustey, Cyril Pitrou, Dominik Schwarz, David Blaschke, and Oleksii Ivanytskyi for the insightful discussions. A.M. and N.C. acknowledge the University of Miami for partial support.
P.N. acknowledges support from the Gordon and Betty Moore Foundation and the John Templeton Foundation, which fund the Black Hole Initiative (BHI) at Harvard University, where she serves as a PI. P.N. also acknowledges support from STScI/NASA via grant JWST-GO-03293024.
\\ \textit{Software:} The analysis has been performed by using the \href{https://www.python.org/}{Python} \citep{10.5555/2011965Python} programming language  and \href{https://jupyter.org/}{Jupyter Notebook} \citep{Kluyver:2016aajupyter} interactive computational environment. Specifically, the following packages have been employed: \href{https://numpy.org/}{Numpy} \citep{harris2020arrayNumpy}, \href{http://roban.github.com/CosmoloPy/}{Cosmolopy} \citep{2020ascl.soft09017Kcosmolopy,Eisenstein1999}, \href{https://matplotlib.org/stable/}{Matplotlib} \citep{Hunter:2007matplotlib}, \href{https://scipy.org/}{Scipy} \citep{2020SciPy-NMeth}, \href{https://pandas.pydata.org/}{Pandas} \citep{the_pandas_development_team_2025_17229934}

This research has made use of data obtained from the Gravitational Wave Open Science Center (gwosc.org), a service of the LIGO Scientific Collaboration, the Virgo Collaboration, and KAGRA. This material is based upon work supported by NSF's LIGO Laboratory which is a major facility fully funded by the National Science Foundation, as well as the Science and Technology Facilities Council (STFC) of the United Kingdom, the Max-Planck-Society (MPS), and the State of Niedersachsen/Germany for support of the construction of Advanced LIGO and construction and operation of the GEO600 detector. Additional support for Advanced LIGO was provided by the Australian Research Council. Virgo is funded, through the European Gravitational Observatory (EGO), by the French Centre National de Recherche Scientifique (CNRS), the Italian Istituto Nazionale di Fisica Nucleare (INFN) and the Dutch Nikhef, with contributions by institutions from Belgium, Germany, Greece, Hungary, Ireland, Japan, Monaco, Poland, Portugal, Spain. KAGRA is supported by Ministry of Education, Culture, Sports, Science and Technology (MEXT), Japan Society for the Promotion of Science (JSPS) in Japan; National Research Foundation (NRF) and Ministry of Science and ICT (MSIT) in Korea; Academia Sinica (AS) and National Science and Technology Council (NSTC) in Taiwan.

\appendix
\section{Matter Power Spectrum}
\label{sec:appendix_PS}
In the case of an extended PBH mass function, \citep{Matteri2025-wc} (their Appendix A, eq. A.5 \footnote{They use a lognormal PBH mass
function (eq. A.1), but we can replace that specific mass function with any extended one.}) give a functional form for the PBH power spectrum:
\begin{equation}
P_{PBH}(k) = \frac{D(z)^2}{\rho_c^2\; \Omega^2_{dm} \ln 10} \int_{0}^{M(k)} {M \frac{d\;n_{PBH}}{d\,\log M} dM}
\end{equation}
where $d\,n_{PBH}/d\,\mathrm{log} M$  is the differential PBH number density, $M$ is the mass of PBHs, $M(k)= 2 \pi^2 \Omega_{dm} {\rho_c}/{k^3}$ is the mass associated with a given critical wavenumber $k$, and $D(z)$ is the growth factor as a function of redshift $z$, which can be expressed as \citep{Inman2019-fy,Liu2022-xk}:
\begin{align}
& D(a) \simeq \left( 1 + \frac{3\gamma}{2a_-} s \right)^{a_-} - 1, 
\qquad s = \frac{a}{a_{\mathrm{eq}}},\\
& \gamma = \frac{\Omega_{\mathrm{m}} - \Omega_{\mathrm{b}}}{\Omega_{\mathrm{m}}},
\qquad 
a_- = \frac{1}{4}\left( \sqrt{1 + 24\gamma} - 1 \right)
\end{align}
where $a = 1/(1+z)$, and $a_{eq} = 1/(1 + z_{eq})$ is the scale factor at matter–radiation equality with $z_{eq} \simeq 3400$.
Moreover, we can relate $d\,n_{PBH}/d\,\mathrm{log} M$ to the PBHs differential fractional abundance $d\,f_{PBH}/d M$ through:
\begin{equation}
\frac{d\;n_{PBH}}{d\,\log M} = \ln 10\;\rho_{dm}\frac{d\,f_{PBH}} {dM} = \ln 10\ \frac{\rho_{dm}}{M}\frac{d\,f_{PBH}} {d\;\ln M} 
\end{equation}
Therefore, the PBH PS can be written as
\begin{align}
P_{PBH}(k) 
&= \frac{D(z)^2 \rho_{dm}}
{\rho_c^2\; \Omega^2_{dm} }
\int_{0}^{M(k)} 
\frac{d\,f_{PBH}}{d\,\ln M}\, dM = \frac{D(z)^2}
{\rho_c\; \Omega_{dm} }
\int_{0}^{M(k)} 
\frac{d\,f_{PBH}}{d\,\ln M}\, dM .
\end{align}

\section{Early merger rate density}
\label{sec:appendix_earlyMRD}
The primordial merger rate derived in \citet{Raidal_2019} assumes that PBH binaries formed during the radiation era evolve in isolation. In practice, interactions with neighbouring PBHs and the growth of small PBH clusters can perturb or disrupt these binaries, reducing the effective merger rate. Following \citet{Raidal_2019}, these effects can be approximately incorporated through multiplicative suppression factors,
\begin{equation}
\tau_E(M_1,M_2,z) \;\rightarrow\;
\tau_E(M_1,M_2,z)\, S_1(M_{\rm tot})\, S_2(z)\, ,
\end{equation}
where $S_1$ accounts for perturbations from nearby PBHs and $S_2$ describes the late-time suppression associated with PBH clustering.

The first factor depends on the expected number of PBHs surrounding the binary and on the moments of the PBH mass distribution. Defining
\begin{equation}
\bar{N}(M_{\rm tot}) =
\frac{M_{\rm tot}}{\langle M \rangle}
\frac{f_{\rm PBH}}{f_{\rm PBH}+\sigma_M},
\end{equation}
where $\sigma_M$ denotes the variance of density perturbations at matter--radiation equality, the survival probability of a primordial binary can be approximated as
\begin{equation}
S_1(M_{\rm tot})
\simeq
1.42
\left[
\frac{ \langle M^2\rangle/{\langle M\rangle^2}}
{\bar{N}(M_{\rm tot}) + C(f_{\rm PBH})}
+ \frac{\sigma_M^2}{f_{\rm PBH}^2}
\right]^{-21/74}
\exp\!\left[-\bar{N}(M_{\rm tot})\right] \, .
\end{equation}
where the function $C(f_{\rm PBH})$ encodes the contribution from Poisson fluctuations in the PBH distribution and can be expressed as:
\begin{equation}
C(f_{\rm PBH}) =
\frac{f_{\rm PBH}^2 \, \langle M^2 \rangle / \langle M \rangle^2}{ \sigma_M^2}
\left[
\left(
\frac{\Gamma(29/37)}{\sqrt{\pi}}\,
U\!\left(\frac{21}{74}, \frac{1}{2}, \frac{5 f_{\rm PBH}^2}{6 \sigma_M^2}\right)
\right)^{-74/21}
- 1
\right]^{-1} \, ,
\end{equation}
where $\Gamma(\cdot)$ is the gamma function and $U(\cdot,\cdot,\cdot)$ denotes the confluent hypergeometric function.

The second suppression factor accounts for the evolution of PBH clusters after matter--radiation equality. Following \citet{Raidal_2019}, it can be written as
\begin{equation}
S_2(z) =
\min\!\left[
1,\,
9.6\times10^{-3}\,
f_{\rm eff}^{-0.65}
\exp\!\left(0.03\,\ln^2 f_{\rm eff}\right)
\right],
\end{equation}
where the effective PBH fraction evolves with cosmic time as
\begin{equation}
f_{\rm eff} =
f_{\rm PBH}\left[\frac{t(z)}{t_0}\right]^{0.44}.
\end{equation}
Together, $S_1$ and $S_2$ provide a phenomenological description of the reduction of the primordial binary merger rate due to dynamical perturbations and clustering of PBHs after their formation.

\section{Halo mass function}
We choose the following functional form to describe the DM halo mass function: 
\begin{align}
\frac{dn}{dM_h}(M_h,z) &= \Phi_{Tinker}(M_h,z) \times C_1(M_h,z)
\end{align}
which is the product of the standard Tinker mass function \citep{Tinker2008-mh} times a correction factor (\cite{Behroozi2013-do}, their Appendix G, Eq. G3), 
\label{sec:appendix_halofunct}
where $C_1(M_h,z)=10^{F1(M_h) \cdot F2(M_h,z)}$ and
\begin{align}
F1(z) &= \frac{0.144}{
1 + \exp\!\left[\,14.79\!\left(\frac{1}{1+z} - 0.213\right)\!\right]
}, \\
F2(M_h, z) &= 
\left(\frac{M_h}{10^{11.5}\,M_\odot}\right)^{
\dfrac{0.5}{
1 + \exp\!\left[\,6.5\!\left(\dfrac{1}{1+z}\right)\!\right]
}}.
\end{align}
The Tinker mass function can be expressed as:
\begin{align}
\Phi_{Tinker}(M_h,z) = f\:(\sigma_M,z)\,
\frac{\rho_m}{M_h}\left|\frac{d\ln \sigma_M^{-1}}{dM}\right| 
\end{align}
where $f(\sigma_M, z)$ is a function parametrized as
\begin{equation}
f(\sigma_M, z) = 
A(z)\,
\left[\,1 + \left(\frac{\sigma_M}{b(z)}\right)^{-a(z)}\right]
\exp\!\left[-\,\frac{c}{\sigma_M^{2}}\right]
\end{equation}
where $A(z),a(z),b(z),c,\alpha$ are parameters calibrated numerically and take the following form:
\begin{align}   
A(z) &= 0.186\,\big(1 + z\big)^{-0.14}, \\
a(z) &= 1.47\,\big(1 + z\big)^{-0.06}, \\
b(z) &= 2.57\,\big(1 + z\big)^{-\alpha} \\
c &= 1.19, \\
\log_{10}\alpha &= 
-\left(\frac{0.75}{\log_{10}\!\left(\dfrac{200}{75}\right)}\right)^{1.2}.
\end{align}
The variance of the linear density field, $\sigma_M^2(z)$, quantifies the rms fluctuation of the linear density field when smoothed over a spherical region enclosing mass $M$ at redshift $z$. It links the linear power spectrum $P(k,z)$ and the Fourier transform of the real-space top-hat window function $W(kR)$ of radius $R$, to halo statistics, and it can be expressed as:
\begin{equation}    
\sigma_M^2(z)
= \frac{1}{2\pi^2}
\int_0^{\infty} k^2\, P_{CDM}(k, z)\, W^2(kR)\, dk,
\end{equation}
\begin{align}
 &\text{where}\,\,\, W(kR)= 3\,\frac{\sin(kR) - kR\cos(kR)}{(kR)^3}\quad\text{and}
\quad R = \left(\frac{3M}{4\pi\,\rho_m}\right)^{1/3}.
\end{align}

\section{Star formation rate density}
\label{sec:appendix_SFRD}
The adopted Star formation Efficiency is the following, and it was chosen from \citep{Harikane2022-rc}:
\begin{align}
\frac{\mathrm{SFR}}{\dot M_h}
&= \frac{2\times 3.2\times 10^{-2}}
{\left(\dfrac{M_h}{\Mchar}\right)^{-1.2} + \left(\dfrac{M_h}{\Mchar}\right)^{0.5}} \times  \Big( 0.53\,\tanh\!\big[\,0.54\,(2.9 - z)\,\big] + 1.53\Big),
\end{align}

The procedure to compute $\dot M_h$  follows from the N-body simulation results in \cite{Behroozi2015-lo} (their Appendix B):
\begin{align}
\begin{split} 
M_{\rm med}(M_0,z) &= M_{13}(z)\,10^{\,f(M_0,z)},\\
M_{13}(z) &= 10^{13.276}\,(1+z)^{3.00}\,\left(1+\frac{z}{2}\right)^{-6.11}\,e^{-0.503\,z},\\
f(M_0,z) &= \log_{10}\!\left(\frac{M_0}{M_{13}(0)}\right)\,
\frac{g(M_0,1)}{g\!\left(M_0,a\right)}, \text{ with } \;a=\frac{1}{1+z}\\
g(M_0,a) &= 1+\exp\!\Big[-4.651\big(a-a_0(M_0)\big)\Big],\\
a_0(M_0)&=0.205-\log_{10}\!\Big(\big(10^{9.649}/M_0\big)^{0.18}+1\Big)
\end{split}
\end{align}
Given $M_h$ and $z$, we can solve $M_{\rm med}(M_0,z)=M_h$ for $M_0$. Then,
\begin{align}
\dot M_{\rm med}(M_h,z) &= \left.\frac{\partial M_{\rm med}(M_0,z)}{\partial z}\right|_{M_0}
\frac{dz}{dt},\;\;\; \text{ where}\;\;\; \frac{dz}{dt}=-(1+z)\,H(z).
\end{align}
so finally the expression for $M_h$ is:
\begin{align}
\dot M_h(M_h,z) &= \dot M_{\rm med}(M_h,z)\times
\Big[0.22\,(a-0.4)^2+0.85\Big]\, \times
\frac{\left(\frac{M_h}{10^{12}\,M_\odot}\right)^{0.05 a}\ }
{\left(1+\frac{10^{11-0.5 a}\,M_\odot}{M_h}\right)^{(0.04+0.1 a)} }
\end{align}

\section{Thompson optical depth}
\label{sec:appendix_thompson}
The evolution of the ionized fraction $x_{ion}$ of the intergalactic medium (IGM) is governed by the \emph{reionization balance equation}:
\begin{equation}
\frac{dx_{ion}}{dt} = \frac{\dot{n}_{\mathrm{ion}}(z)}{\bar{n}_{\mathrm{H},0}} - \frac{x_{ion}}{t_{\mathrm{rec}}(z)} ,
\end{equation}
\noindent
In this expression, $\dot{n}_{\mathrm{ion}}(z)$ is the comoving production rate of hydrogen-ionizing photons. 
It is set by the star formation rate density $\rho_{\mathrm{SFR}}(z)$, the escape fraction of Lyman-continuum photons $f_{\mathrm{esc}}$, 
and the ionizing photon production efficiency $\xi_{\mathrm{ion}}$ through the following equation:
\begin{equation}
\dot{n}_{\mathrm{ion}}(z) = f_{\mathrm{esc}} \, \xi_{\mathrm{ion}} \, \rho_{\mathrm{SFR}}(z).
\end{equation}
We investigated the SFRD as given by Eq. \ref{eq:SFRD} for 
the $\Lambda$CDM and the $\Lambda$CDM+PBH cases, setting  $\xi_{\mathrm{ion}}= 10^{53.35}\;\;\text{photons}\;\;s^{-1} M_\odot^{-1} yr $ \citep{Topping_2015} and $f_{esc}=0.3$.
Moreover, $\bar{n}_{\mathrm{H},0}$ is the mean comoving hydrogen number density today, obtained from the primordial hydrogen fraction $X_p = 0.7547$,  
the baryon density parameter $\Omega_b$, the critical density $\rho_c$, and the proton mass $m_p$. 
This sets the normalization against which the ionizing photon budget is compared.

\begin{equation}
\bar{n}_{\mathrm{H},0} = \frac{X_p \, \Omega_b \, \rho_c}{m_p} ,
\end{equation}
Finally, the following relation defines the (inverse of the) recombination time: shorter $t_{\mathrm{rec}}$ means that the IGM recombines more efficiently, 
slowing the progress of reionization. It depends on the clumping of ionized gas ($C_{\mathrm{HII}}$), 
the case B recombination coefficient $\alpha_B(T) = 2.59 \times 10^{-13}\, T_4^{-0.845}
\;\; \mathrm{cm^3\,s^{-1}} $, scaled to an IGM temperature $T = (10^4 K)T_4$ \citep{Michael_Shull2012-bo} and evaluated at $T=2000K$, and the increasing density of the Universe through $(1+z)^3$, 
with helium adding extra electrons via the factor $\left(1 + \tfrac{Y_p}{4X_p}\right)$, where $Y_p = 1 - X_p = 0.2453$.

\begin{equation}
t_{\mathrm{rec}}^{-1}(z) = C_{\mathrm{HII}} \, \alpha_B(T) \, \bar{n}_{\mathrm{H},0} \, (1+z)^3 \, \left(1 + \frac{Y_p}{4X_p}\right) .
\end{equation}
The Thomson optical depth is then defined as the integral of the electron density along the line of sight:
\begin{equation}
\tau(z) = c \, \sigma_T \int_{0}^{z} dz' \, \frac{n_e(z')}{(1+z')H(z')} ,
\end{equation}
where $c$ is the speed of light, $\sigma_T$ is the Thomson scattering cross section, $n_e(z)$ is the proper number density of free electrons, and $H(z) = H_0 \, \sqrt{\Omega_m (1+z)^3 + \Omega_\Lambda}$ is the expansion rate of the Universe.

The electron density can be written as:
\begin{equation}
n_e(z) = f_e(z) \, x_{ion}(z) \, \bar{n}_{\mathrm{H},0} \, (1+z)^3 ,
\end{equation}
where $f_e(z)$ is the number of free electrons per hydrogen nucleus.  
A compact expression for $f_e(z)$ is obtained by introducing a parameter $\eta = \eta(z)$:
\begin{equation}
f_e(z) = 1 + \eta(z) \, \frac{Y_p}{4X_p} ,
\end{equation}
with
\begin{equation}
\eta(z) =
\begin{cases}
1, & z > 4 \quad (\text{He\,II}) \\[6pt]
2, & z \leq 4 \quad (\text{He\,III})
\end{cases}
\end{equation}
Here $\eta$ is a parameter that encodes the helium ionization state: 
$\eta = 1$ corresponds to singly ionized helium (He\,II, dominant at $z>4$), 
while $\eta = 2$ corresponds to doubly ionized helium (He\,III, dominant at $z \leq 4$).
Substituting $n_e(z)$ into the expression for $\tau(z)$ gives:
\begin{equation}
\tau(z) = c \, \sigma_T \, \bar{n}_{\mathrm{H},0} \int_{0}^{z} dz' \, 
\frac{f_e(z') \, x_{ion}(z') \, (1+z')^2}{H(z')} .
\end{equation}

\bibliography{sample701}{}
\bibliographystyle{aasjournalv7}

\end{document}